\documentclass[11pt,xcolor=dvipsnames]{article}
\pdfoutput=1 

\DeclareMathAlphabet{\scr}{U}{rsfs}{m}{n}

\usepackage{scalerel,stackengine}
\usepackage[colorlinks,urlcolor=black,linkcolor = black,citecolor = black,]{hyperref}
\usepackage{subfigure}
\usepackage{latexsym}
\usepackage{epsfig}
\usepackage[mathscr]{eucal}
\usepackage{amsfonts}
\usepackage{amscd}
\usepackage{cite}
\usepackage{array}
\usepackage{amssymb}
\usepackage{colordvi}
\usepackage[centertags]{amsmath}
\usepackage{enumerate}
\usepackage{graphicx}
\usepackage{booktabs}
\usepackage{theorem}
\usepackage[footnotesize]{caption}
\usepackage{soul}
\usepackage{slashed}
\usepackage[obeyFinal]{todonotes}
\usepackage{xcolor}
\usepackage{bbm}
\usepackage[utf8]{inputenc}
\usepackage{fancyvrb}
\usepackage{framed}
\usepackage{xspace}
\usepackage{braket}
\usepackage{cancel}
\usepackage{nicefrac}

\usepackage{cleveref}
\crefname{chapter}{Chapter}{Chapter}
\crefname{section}{Sec.}{Secs.}
\crefname{table}{Tab.}{Tabs.}
\crefname{figure}{Fig.}{Figs.}
\crefname{equation}{Eq.}{Eqs.}
\crefname{appendix}{Appendix\ }{Appendix\ }
\Crefname{section}{Section}{Sections}
\Crefname{table}{Tale}{Tables}
\Crefname{figure}{Figure}{Figures}
\Crefname{equation}{Equation}{Equations}

\newcommand{\hc}{\text{ h.c.}}

\newcommand{\gev}{~\text{GeV}}

\newcommand{\lsim}{\raisebox{-0.13cm}{~\shortstack{$<$ \\[-0.07cm]
      $\sim$}}~}

\newcommand{\s}{\newline \vspace*{-3.5mm}}

\newcommand{\beq}{\begin{eqnarray}}
\newcommand{\eeq}{\end{eqnarray}}

\newcommand{\ii}{\mathrm{i}~}

\newcommand{\Tr}{\mathrm{Tr}}

\newcommand{\e}{\text{e}}
\newcommand{\Lio}{\textbf{ L}}

\newcommand{\figref}[2][{}]{\hyperref[#2]{\figurename~\ref{#2}#1}} 

\newenvironment{kasten*}[1]
{
\hspace{0.05\linewidth}
\begin{minipage}{0.95\linewidth}
\setlength{\fboxsep}{10pt}
\definecolor{shadecolor}{gray}{0.9}
\definecolor{framecolor}{gray}{0}

\MakeFramed {\FrameRestore}
\subsection*{#1}
}
{
\endMakeFramed
\end{minipage}
\vspace{1em}
}

\newcommand{\hup}{h_{\uparrow}}
\newcommand{\hdown}{h_{\downarrow}}
\newcommand{\mdown}{m_{h_{\downarrow}}}
\newcommand{\mup}{m_{h_{\uparrow}}}
\newcommand{\mHc}{m_{H^{\pm}}}
\newcommand{\vw}{v_W}

\newcommand{\ewbgBSMPT}{\text{\tt BSMPT v2}\xspace}

\newcommand{\FH}{{\tt FH}\xspace}
\newcommand{\VIA}{{\tt VIA}\xspace}
\newcommand{\ScannerS}{{\tt ScannerS}\xspace}
\newcommand{\cbrak}[1]{\left(#1\right)}

\newcommand{\wCP}{\omega_{\text{CP}}}

\newcommand{\wone}{\omega_1}
\newcommand{\wtwo}{\omega_2}
\newcommand{\primed}[1]{#1^{\prime}}

\allowdisplaybreaks

\newcommand{\omegaCB}{\ensuremath{\omega_{\text{CB}}}}
\newcommand{\omegaCP}{\ensuremath{\omega_{\text{CP}}}}

\begin{document}
\title{
	\vspace*{-3cm}
	\phantom{h} \hfill\mbox{\small KA-TP-14-2021}
	\\[1cm]
	\textbf{
		Electroweak Baryogenesis in the \\CP-Violating Two-Higgs Doublet Model\\[4mm]}}

\date{}
\author{
Philipp Basler\footnote{E-mail:
	\texttt{philipp.basler@alumni.kit.edu}} ,
Margarete M\"{u}hlleitner$^{1\,}$\footnote{E-mail:
	\texttt{margarete.muehlleitner@kit.edu}} ,
Jonas M\"{u}ller $^{1\,}$\footnote{E-mail: \texttt{jonas.mueller@kit.edu}}
\\[9mm]
{\small\it
$^1$Institute for Theoretical Physics, Karlsruhe Institute of Technology,} \\
{\small\it 76128 Karlsruhe, Germany}}

\maketitle

\begin{abstract}
Recently we presented the upgrade of our code {\tt BSMPT} for the
calculation of the electroweak phase transition (EWPT) to  {\tt BSMPT
  v2} which now includes the computation of the baryon asymmetry of
the universe (BAU) in the CP-violating 2-Higgs-Doublet Model
(C2HDM). In this paper we use {\tt BSMPT v2} to investigate the size of the
BAU that is obtained in the C2HDM with the two implemented approaches {\tt FH} and
{\tt VIA} to derive the transport equations, by taking into account
all relevant theoretical and experimental constraints. We identify 
similarities and differences in the results computed with the two
methods. In particular, we analyse the dependence of the obtained BAU
on the parameters relevant for successful baryogenesis. Our
investigations allow us to pinpoint future directions for improvements
both in the computation of the BAU and in possible avenues taken for
model building. 
\end{abstract}
\thispagestyle{empty}
\vfill
\newpage
\setcounter{page}{1}

\section{Introduction}
The Standard Model (SM) of particle physics has seen a great success
with the commissioning of the Large Hadron Collider (LHC) where the
last missing piece of the SM, the Higgs boson, was discovered in
2012 by the LHC experiments ATLAS \cite{Aad:2012tfa} and CMS
\cite{Chatrchyan:2012xdj}. The success story is clouded by the fact
that there are remaining puzzles that cannot be explained within the
SM. While the discovered Higgs boson behaves very SM-like
\cite{Aad:2015mxa,Khachatryan:2014kca,Aad:2015gba,Khachatryan:2014jba}
open questions like the observed baryon-antibaryon asymmetry $\eta$ of the
universe \cite{Bennett:2012zja} call for new physics extensions. The
asymmetry can be generated dynamically through electroweak
baryogenesis
\cite{Kuzmin:1985mm,Cohen:1990it,Cohen:1993nk,Quiros:1994dr,Rubakov:1996vz,Funakubo:1996dw,Trodden:1998ym,Bernreuther:2002uj,Morrissey:2012db}
provided the three Sakharov conditions \cite{Sakharov:1967dj} are
fulfilled. These are baryon number violation, C and CP violation and
departure from the thermal equilibrium. The mechanism requires
the electroweak phase transition (EWPT), which proceeds through bubble
formation, to be of strong first order
\cite{Trodden:1998ym,Morrissey:2012db} so that the baryon number violating sphaleron
transitions in the false vacuum \cite{Manton:1983nd,Klinkhamer:1984di}
can be suppressed. Although all three Sakharov conditions are 
in principle met by the SM a strong first order EWPT (SFOEWPT) can only be
realized for an SM Higgs boson mass around 70-80 GeV \cite{Aad:2015zhl}
which is in contradiction with the experimentally measured 125~GeV
\cite{Aad:2015zhl}. Additionally, the amount of CP violation in the SM
that stems from the Cabibbo-Kobayashi-Maskawa (CKM) matrix is not large enough
to quantitatively reproduce the measured value of $\eta$
\cite{Morrissey:2012db,Gavela:1993ts}. These problems can be remedied by
extended Higgs sectors that provide additional sources of CP violation
and further scalar bosons that may trigger an SFOEWPT. An example is
the 2-Higgs-Doublet Model (2HDM) where an SFOEWPT can be realized in
accordance with all relevant theoretical and experimental constraints
both for a CP-conserving
\cite{previous2hdm,Basler:2016obg,Laine:2017hdk,Dorsch:2017nza,Andersen:2017ika,Bernon:2017jgv,Wang:2018hnw,Kainulainen:2019kyp}
and a CP-violating Higgs potential
\cite{previousc2hdm,C2HDM,Wang:2019pet}.
 \s

Denoting by $v_c$ the vacuum expectation value (VEV) at the critical
temperature $T_c$ which is defined as the temperature where two
degenerate global minima exist, a value of $v_c/T_c$ larger than one
is indicative for an SFOEWPT \cite{Quiros:1994dr,Moore:1998swa}.
There are several programs on the marked that allow for the
computation of the minima of extended Higgs sectors\footnote{The {\tt
    C++} library {\tt EVADE} \cite{Hollik:2018wrr,Ferreira:2019iqb}
  studies the vacuum stability at zero 
  temperature, but can be applied to beyond-the-SM (BSM) models with many scalar
  fields while at the same time being fast and efficient.} and the strength
of the phase transition. Thus {\tt 
  Vevacious} \cite{Camargo-Molina:2013qva,Camargo-Molina:2014pwa}
allows to find the global minima of the one-loop effective potential with
many scalars.
{\tt CosmoTransitions} \cite{Wainwright:2011kj} is a tool that analyzes the different vacua of
theories with single or multiple scalar fields in order to determine
the critical temperatures, the super-cooling temperatures and the
bubble wall profiles separating the different phases and that also
describes their tunneling dynamics. The {\tt C++} package {\tt
  PhaseTracer} \cite{Athron:2020sbe} finds the cosmological phases and the critical
temperatures for phase transitions for any scalar potential by tracing
the minima of the effective potential as function of the temperature
change. 
We have published the {\tt C++} code {\tt BSMPT} \cite{Basler:2018cwe,
  Basler:2020nrq} for the 
calculation of the strength of the EWPT of extended Higgs sectors by determining the critical
VEV $v_c$ at the temperature $T_c$. For this we compute the loop-corrected
effective potential at finite temperature
\cite{Coleman:1973jx,Quiros:1999jp,gauge1} including the daisy resummation
of the bosonic masses \cite{Carrington:1991hz}. For efficient
parameter scans in the models under investigation we use a
renormalization scheme that keeps the loop-corrected Higgs masses and
mixing angles at their tree-level values and applied it to investigations
in the 2HDM \cite{Basler:2016obg}, the
C2HDM \cite{C2HDM,Basler:2019iuu}
and the Next-to-2HDM (N2HDM) \cite{Basler:2019iuu}. \s

While the SFOEWPT is a necessary
condition for successful baryogenesis the question still
remains to be answered if the model is able to generate the observed baryon asymmetry
of the universe after taking into account all relevant constraints.
The baryon asymmetry that can be estimated by assuming that all cosmic
microwave background photons are the result of thermal annihilation of
the baryon pairs is given by \cite{ParticleDataGroup:2018ovx}
\beq
\frac{n_B - n_{\bar{B}}}{n_B + n_{\bar{B}}} \approx \eta \equiv \frac{n_B}{n_\gamma} = (6.2 \pm 0.4) \cdot 10^{-10} \;,
\eeq
where $n_B$ ($n_{\bar{B}}$) is the (anti-)baryon density and
$n_\gamma$ the photon density of today's universe. Recently, we
published the upgrade {\tt BSMPT v2} \cite{Basler:2020nrq}. It not
only extends the existing code 
to include the computation of the strength of the EWPT for the already
implemented models (the 2HDM, C2HDM and the N2HDM) by the complex
singlet extension of the SM (CxSM). But its 
major upgrade is the implementation of the computation of the baryon
asymmetry of the universe for the C2HDM in two different
approximations, the so-called {\tt FH}
\cite{Cline1998,Kainulainen2001,Fromme:2006wx,Fromme:2006cm} and the
{\tt VIA} approach \cite{Riotto:1995hh, Riotto:1997vy,
  Lee:2004we,Chung:2009qs}. In that paper, we explained 
our treatment of the wall profile and described in detail the two approaches for the
computation of the BAU, as well as the code, the system requirements,
the installation and the usage of the program. We complemented the manual
by a more general discussion on the approaches and approximations that
we used. \s

The purpose of this work now is to apply our code {\tt BSMPT v2} to
the computation of the BAU in the C2HDM. We want to investigate how
the two different approaches used in the literature compare to each
other.\footnote{The BAU in the Minimal Supersymmetric Extension of
the Standard Model (MSSM) has been calculated with the {\tt VIA} method
in \cite{Carena:1997gx} {\it e.g.}, and with the {\tt FH} approach in
\cite{Cline:1997vk,Cline:2000nw,Cline:2000kb}. A short general
comparison of the derivation of the quantum transport
equations from first principles in the Schwinger-Keldysh 
formalism with the {\tt FH} and {\tt VIA} approach is presented in
\cite{Konstandin:2013caa} as well as a quantitative comparison between
the different approaches applied to the MSSM. In \cite{Cline:2020jre},
a comparison was performed for a prototypical model of CP violation in
the wall.} While the approaches lead to quite different results we
will analyse where they show similar behaviour and what are the crucial
parameters that influence the possible size of $\eta$. Our goal is to
find out if in principle it is possible to obtain a BAU in the
C2HDM that is compatible with the observed value while at the same
time applying the relevant theoretical and experimental
constraints. We furthermore investigate the inclusion of different
fermion species, top, bottom and $\tau$, in the {\tt VIA} approach and their
impact on the BAU. The paper will give us more insights in the effects of
the various approaches used to compute the BAU and will pinpoint
different directions to be taken both for future model building and
for improvement of the computation. \s

The paper is organized as follows. In Sec.~\ref{sec:c2hdm} we 
introduce the C2HDM and set our notation. We briefly comment on the
computation of the EWPT and take the occasion to present the
improvement of our renormalization scheme of the C2HDM implemented in
{\tt BSMPT v2.2}. Section \ref{sec:baryocalc} is devoted to the
calculation of the BAU. We discuss in detail our calculation of the
bubble wall profile and summarize the computation of the BAU in the
{\tt FH} and in the {\tt VIA} approach, an extensive description is
given in \cite{Basler:2020nrq}. Section~\ref{sec:numerical} contains
our numerical analysis. After the description of the applied
constraints and our parameter scan, we discuss the newly implemented
counterterms in the C2HDM before moving on to the presentation of the
results on the outcome of the BAU in the two applied approaches, and
the dependence on the bubble wall velocity. We investigate the interplay between
wall thickness and overall mass scale of the Higgs spectrum and
analyse the behaviour of both approaches with respect to the
parameters that are crucial for successful BAU. Finally, we present
the effect of additional fermions included in the {\tt VIA}
computation. We conclude in Sec.~\ref{sec:concl}.  

\section{The Complex Two-Higgs Doublet Model \label{sec:c2hdm}}
We give a brief introduction in the scalar sector of the
C2HDM \cite{Branco:1985aq,Ginzburg:2002wt,Khater:2003wq} and
refer for a more detailed discussion of the model to 
\cite{C2HDM,Fontes:2017zfn}. The scalar potential of the C2HDM is a
simple extension of the SM Higgs sector with an additional
$\mathrm{SU}(2)$ Higgs doublet
\begin{align}
	V_{\text{C2HDM}} =~ & m_{11}^2 \Phi_1^{\dagger}\Phi_1 + m_{22}^2
	\Phi_2^{\dagger}\Phi_2 + \frac{\lambda_1}{2}
	\left(\Phi_1^{\dagger}\Phi_1\right)^2 +
	\frac{\lambda_2}{2}
	\left(\Phi_1^{\dagger}\Phi_1\right)^2+ \lambda_3
	\left(\Phi_1^{\dagger}\Phi_1\right)\left(
	\Phi_2^{\dagger}\Phi_2\right) \nonumber\\
	                    & + \lambda_4 \left( \Phi_1^\dagger \Phi_2\right)^2  + \left[
		\frac{\lambda_5}{2} \left(\Phi_1^\dagger\Phi_2\right)^2 - m_{12}^2
		\left(\Phi_1^\dagger \Phi_2\right) + h.c. \right] \,, \label{2HDM:Pot}
\end{align}
with a softly broken discrete $\mathbb Z_2$ symmetry under which
$\Phi_1\rightarrow \Phi_1$ and $\Phi_2\rightarrow -\Phi_2$. This
$\mathbb Z_2$ symmetry ensures the absence of flavour-changing neutral
currents (FCNC) at tree level and allows for different types of the
C2HDM depending on how the Higgs doublets couple to the fermions. The
different possibilities are listed in 
\cref{tab::TypeTable}. For simplicity only Type I and II are discussed
in this analysis.
\begin{table}[b]
	\center
	\begin{tabular}{c c c c | c c c c c }
		\toprule
		                & $u$-type & $d$-type & leptons  & Q & $u_R$ & $d_R$ & L & $l_R$ \\\midrule
		Type I          & $\Phi_2$ & $\Phi_2$ & $\Phi_2$ & + & $-$   & $-$   & + & $-$   \\
		Type II         & $\Phi_2$ & $\Phi_1$ & $\Phi_1$ & + & $-$   & +     & + & $-$   \\
		lepton-specific & $\Phi_2$ & $\Phi_2$ & $\Phi_1$ & + & $-$   & +     & + & $-$   \\
		flipped         & $\Phi_2$ & $\Phi_1$ & $\Phi_2$ & + & $-$   & $-$   & + & +     \\\bottomrule
	\end{tabular}
	\caption{Left: Definition of the 2HDM types through the allowed couplings between
		fermions and Higgs doublets. Right: Corresponding $\mathbb{Z}_2$
		parity assignments to the left-handed quark and lepton doublets, $Q$,
		$L$, and the right-handed singlets of the up-type and down-type
		quarks, $u_R$ and $d_R$, and right-handed leptons $l_R$.}
	\label{tab::TypeTable}
\end{table}
All Lagrangian parameters are real due to the hermiticity of the
potential except for $m_{12}^2$ and $\lambda_5$ which can be complex
as we allow for CP violation. Upon electroweak
symmetry breaking (EWSB) the two Higgs doublets acquire vacuum
expectation values (VEVs) around which they can be expanded in terms of
the charged CP-even and CP-odd field components $\rho_i$ and $\eta_i$
and the neutral CP-even and CP-odd fields $\zeta_i$ and
$\psi_i$ ($i=1,2$). The general tree-level vacuum structure of the
2HDM allows for three different possible vacua, the normal EW-breaking
vacuum, a CP-breaking and a charge-breaking (CB) vacuum. As was shown in
Ref.~\cite{Ferreira:2004yd, Barroso:2007rr,Ivanov:2007de} these vacua cannot
coexist simultaneously at tree level. Higher-order corrections or
finite temperature effects might break this statement, hence we allow
for a more general vacuum structure in the analysis.  Denoting the
corresponding VEVs by $\omega_{1,2}$ for the normal vacuum, and by
$\omega_{\text{CP}}$ and $\omega_{\text{CB}}$ for the CP-breaking and
the charge-breaking minimum, respectively, the expansion of the two
Higgs doublets $\Phi_i$ around the VEVs is given by
\begin{align}
	\Phi_1 = & \frac{1}{\sqrt{2}} \begin{pmatrix}
		\rho_1 + \ii \eta_1 \\ \zeta_1 + \omega_1 + \ii \psi_1
	\end{pmatrix} \qquad \Phi_2 = \frac{1}{\sqrt{2}} \begin{pmatrix}
		\rho_2 + \omegaCB +\ii \eta_2   \\ \zeta_2 + \omega_2 + \ii
		\left(\psi_2 + \omegaCP \right)
	\end{pmatrix}  \;,
\end{align}
with
\begin{align}
	\braket{\Phi_1} = & \frac{1}{\sqrt{2}}\begin{pmatrix}
		0 \\ \omega_1
	\end{pmatrix}	\quad \mbox{and} \quad \braket{\Phi_2} =
	\frac{1}{\sqrt{2}} \begin{pmatrix}
		\omegaCB \\ \omega_2 + \ii \omegaCP
	\end{pmatrix} \,,
	\label{eq:2hdmdoublets1}
\end{align}
where the bracket $\braket{\dots}$ indicates the vacuum state. The vacuum
structure at zero temperature is denoted as
\begin{equation}
	v_i \equiv \omega_i\big\vert_{T=0}\,  \qquad i =
	1,2,\mathrm{CP},\mathrm{CB} \;,
	\label{eq:2hdmdoublets2}
\end{equation}
with
\begin{equation}
	v_{\text{CP}}=v_{\text{CB}} \equiv 0\,.
\end{equation}
This ensures that we end up in the physical
minimum given by the normal EW tree-level minimum at zero temperature. A non-zero value
for the CB VEV would break electric 
charge conservation and introduce massive photons. Therefore all
parameter points showing such unphysical vacuum structures are
neglected in the analysis as well as those breaking CP
  invariance. The VEVs of the normal EW minimum are related to the SM VEV by
\begin{equation}
	v_1^2+v_2^2 \equiv v^2 \approx \left( 246\gev\right)^2\,.
\end{equation}
The minimum conditions of the potential read
\begin{equation}
	\frac{\partial V_{\text{tree}}}{\partial\Phi_i^{\dagger}}\big\vert_{\Phi_j=\braket{\Phi_j}} \overset{!}{=} 0 \,,\quad   i,j\in\lbrace 1,2\rbrace\,,
\end{equation}
where $\braket{\Phi_j}=(0,v_j/\sqrt{2})^T$ at $T=0$ lead to the tadpole
conditions
\begin{subequations}
	\label{tadpole::all}
	\begin{align}
		m_{11}^2 = \mbox{Re}~m_{12}^2 \frac{v_2}{v_1} - \frac{\lambda_1}{2} v_1^2 - \frac{\lambda_3+\lambda_4\mbox{Re}\lambda_5}{2} v_2^2 \label{C2HDM::tadpol1} \\
		m_{22}^2 = \mbox{Re} m_{12}^2 \frac{v_1}{v_2} - \frac{\lambda_2}{2} v_2^2 - \frac{\lambda_3+\lambda_4+\mbox{Re}\lambda_5}{2} v_1^2 \label{C2HDM::tadpol2} \\
		\mbox{Im} m_{12}^2 = \mbox{Im} \lambda_5 \frac{v_1v_2}{2}\label{C2HDM::tadpol3} \,,
	\end{align}
\end{subequations}
which allow us to trade the Lagrangian parameters $m_{11}^2$, $m_{22}^2$ for
the zero-temperature EW VEVs $v_1$ and $v_2$. \Cref{C2HDM::tadpol3}
relates the two phases of the complex parameters $m_{12}^2$ and
$\lambda_5$ and we follow the conventions defined in
\cite{Fontes:2017zfn}. \s

The mass eigenstates of the charged sector, the charged Higgs bosons
$H^\pm$ and the charged Goldstone bosons $G^\pm$, are obtained through
the rotation
\begin{align}
	\begin{pmatrix}
		G^\pm \\ H^\pm
	\end{pmatrix} = & R_\beta \begin{pmatrix}
		\frac{1}{\sqrt{2}} \left( \rho_1 \pm \ii \eta_1 \right) \\ \frac{1}{\sqrt{2}} \left( \rho_2 \pm\ii \eta_2 \right)
	\end{pmatrix}
\end{align}
with the rotation matrix
\begin{align}
	R_\beta = & \begin{pmatrix}
		\cos(\beta) & \sin(\beta) \\ -\sin(\beta) & \cos(\beta)
	\end{pmatrix}
\end{align}
and the mixing angle $\beta$ defined through
\begin{align}
	\tan\beta = & \frac{v_2}{v_1} \,.
\end{align}
Applying the same rotation matrix to the CP-odd fields yields the
neutral Goldstone boson $G^0$ and the CP-odd field $\zeta_3$ as
\begin{align}
	\begin{pmatrix}
		G^0 \\ \zeta_3
	\end{pmatrix} = & R_\beta \begin{pmatrix}
		\psi_1 \\ \psi_2
	\end{pmatrix} \,.
\end{align}
The mass eigenstates of the neutral Higgs sector, $H_k~(k=1,2,3)$, are then given by
\begin{equation}
	\label{C2HDM::rotation}
	\begin{pmatrix}
		H_1 \\H_2\\H_3
	\end{pmatrix}
	=R \begin{pmatrix} \zeta_1\\\zeta_2\\\zeta_3 \end{pmatrix} \,,
\end{equation}
with the rotation matrix ($ c_i \equiv\cos\alpha_i,$ $s_i \equiv
\sin\alpha_i$, $i=1,2,3$)
\begin{equation}
	R = \begin{pmatrix}
		c_1c_2                            & s_1 c_2                         & s_s     \\
		-\left(c_1 s_2s_3+ s_1 c_3\right) & c_1 c_3 - s_1 s_2 s_3           & c_2 s_3 \\
		-c_1s_2c_3+s_1 s_3                & -\left(c_1 s_3+s_1s_2c_3\right) & c_2 c_3
	\end{pmatrix}\,.
	\label{eq:RotationMatrix}
\end{equation}
Without loss of generality the mixing angles $\alpha_i$ can be chosen in the interval
\begin{equation}
	-\frac{\pi}{2}\leq \alpha_i < \frac{\pi}{2}\,.
\end{equation}
The rotation \cref{C2HDM::rotation} yields a diagonal mass matrix
\begin{equation}
	R M^2_{\text{Scalar}} R^T = \operatorname{diag}(m_{H_1}^2,m_{H_2}^2,m_{H_3}^2)\,,
	\label{scalar_mass_matrix}
\end{equation}
with mass ordered neutral Higgs boson masses
\begin{equation}
	m_{H_1}\leq m_{H_2}\leq m_{H_3}\,.
\end{equation}
The C2HDM potential can then be expressed in terms of the following
nine independent input parameters
\begin{equation}
	v\,,\quad \tan\beta\,,\quad \alpha_{1,2,3}\,,\quad m_{H_i}\,,\quad
	m_{H_j}\,,\quad m_{H^{\pm}}\,\quad \text{and}\,\quad \mbox{Re}(m_{12}^2)\,.
\end{equation}
Here, $m_{H_i}$ and $m_{H_j}$ denote any two of the three neutral
Higgs bosons, with one of them being the 125 GeV scalar. The remaining
mass is expressed in terms of the other two Higgs boson masses and
elements of the rotation matrix defined in \cref{eq:RotationMatrix}
through the relation \cite{ElKaffas:2007rq}
\begin{align}
	\sum\limits_{k=1}^{3} m_{H_k}^2 R_{k3} \left( R_{k2} \tan\beta - R_{k1} \right) = 0 \,,
\end{align}
so that it is no direct input parameter in our parameter scan.

\subsection{Computation of the Phase Transition}
In \cite{Basler:2016obg,C2HDM,Basler:2018cwe} we presented in detail
the computation of the loop-corrected effective potential at finite
temperature from which we deduce the critical VEV $v_c$ at the
critical temperature $T_c$, which denotes the temperature where the
symmetric and non-symmetric vacuum become degenerate. For values of
$\xi_c = v_c/T_c \ge 1$ we have a strong first order EWPT
\cite{Quiros:1994dr,Moore:1998swa}. We have chosen the renormalization
conditions of the loop-corrected effective potential such that not
only the VEV and all physical Higgs boson masses, but also all mixing
matrix elements remain at their tree-level values. This choice allows
us to effectively determine in a parameter scan of the model parameter
points that are compatible with the theoretical and experimental
constraints without the need to resort to an iterative procedure as we
can directly use the tree-level mass values and mixings as input
parameters. In the CP-violating 2HDM, however, the determination of
the counterterm potential from the parametrization of the tree-level
potential is not sufficient to render all masses and mixing values
equal to their tree-level values. At one-loop level new
flavour-violating structures are induced due to CP violation. This has
to be taken into account in the construction of the counterterm
potential which is hence given by
\beq
V_{\text{CT}} &=& \frac{\delta m_{11}^2}{2} \omega_1^2 + \frac{\delta
m_{22}^2}{2} (\omega_2^2+\omega_{\text{CP}}^2 + \omega_{\text{CB}}^2) 
- \delta \mbox{Re}(m_{12}^2) \,\omega_1\omega_2 + \delta
\mbox{Im}(m_{12}^2) \,\omega_1\omega_{\text{CP}} + \frac{\delta \lambda_1}{8}
\omega_1^4 \nonumber \\
&+& \frac{\delta \lambda_2}{8} (\omega_2^2+\omega_{\text{CP}}^2+\omega_{\text{CB}}^2)^2
+\frac{\delta\lambda_3}{4} \omega_1^2\left(\omega_{2}^2
  +\omega_{\text{CP}}^2 + \omega_{\text{CB}}^2\right) 
+ \frac{\delta\lambda_4}{4}
\omega_1^2\left(\omega_2^2+\omega_{\text{CP}}^2\right)  \nonumber\\
&+& \frac{\delta \mbox{Re}(\lambda_5)}{4} \omega_1^2
\left(\omega_2^2-\omega_{\text{CP}}^2\right) -
\frac{\delta\mbox{Im}(\lambda_5)}{2}
\omega_1^2\omega_2\omega_{\text{CP}} \nonumber \\
&+& \delta T_1 \, \omega_1 + \delta T_2 \, \omega_2
+ \delta T_{\text{CP}} \, \omega_{\text{CP}} \nonumber \\
&-& \frac{\delta \mbox{Im}(\lambda_6)}{2} \omega_1^2
\omega_{\text{CP}}
- \frac{\delta \mbox{Im}(\lambda_7)}{2} \omega_2^2 \omega_{\text{CP}}
\;. \label{eq:countertermpot}
\eeq
This form of the counterterm potential differs by the last two terms
from the one given in Ref.~\cite{C2HDM}. A check of the results given
in \cite{C2HDM} shows, however, that the difference induced by the two
new terms is negligible so that the results given in \cite{C2HDM} do
not change significantly. \s

We apply the following renormalization conditions \cite{C2HDM} 
\beq
\partial_{\phi_i} \left.V_{\text{CT}} (\phi)\right|_{\phi= \langle
  \phi^c \rangle_{T=0}} &=&
- \partial_{\phi_i} \left.V_{\text{CW}} (\phi) \right|_{\phi=
  \langle\phi^c \rangle_{T=0}} \label{eq:rencond1} \\
\partial_{\phi_i} \partial_{\phi_j}\left.V_{\text{CT}}
  (\phi)\right|_{\phi= \langle\phi^c\rangle_{T=0}} &=&
- \partial_{\phi_i} \partial_{\phi_j}\left.V_{\text{CW}} (\phi)\right|_{\phi=
  \langle\phi^c\rangle_{T=0}} \label{eq:rencond2} \;,
\eeq
with the Coleman-Weinberg potential $V_{\text{CW}}$ given in \cite{C2HDM}, 
\beq
\phi_i \equiv \{ \rho_1, \eta_1, \rho_2, \eta_2, \zeta_1, \psi_1,
\zeta_2, \psi_2 \} \;,
\eeq
and the field configuration $\langle \phi^c \rangle_{T=0}$ in
the minimum at $T=0$,
\beq
\langle \phi^c \rangle_{T=0} = (0,0,0,0,v_1,0,v_2,0) \;.
\eeq
These conditions ensure the EW minimum to be a local minimum at $T=0$,
which we check numerically to be the global one, and that the masses
and mixing angles remain at their tree-level values at $T=0$. Since
the conditions are not enough to fix all renormalization constants, we
have to choose two of them and set them equal to $t_1$ and $t_2 \in
\mathbb{R}$, respectively. This results in the following counterterms
in terms of the derivatives of the potential,
\begin{subequations}
  \begin{align}
    \delta m_{11}^2 &= t_1 v_2^2 - \frac{3}{2} H^\text{CW}_{\psi_1,
                      \psi_1} - \frac{v_2 }{2 v_1}
                      H^\text{CW}_{\psi_1, \psi_2}+ \frac{1}{2}
                      H^\text{CW}_{\zeta_1, \zeta_1} + \frac{v_2 }{2
                      v_1}H^\text{CW}_{\zeta_1, \zeta_2} \label{eq:glgb}\\
    \delta m_{22}^2 &= t_1 v_1^2 - \frac{v_1 }{2 v_2}H^\text{CW}_{\psi_1, \psi_2} - \frac{3}{2} H^\text{CW}_{\psi_2, \psi_2} + \frac{v_1 }{2 v_2}H^\text{CW}_{\zeta_1, \zeta_2} + \frac{1}{2} H^\text{CW}_{\zeta_2, \zeta_2}\\
    \delta \mathrm{Im}\left(m_{12}^2\right) &= \frac{1}{2} H^\text{CW}_{\zeta_1, \psi_2} + \frac{v_2 }{2 v_1}H^\text{CW}_{\zeta_2, \psi_2} + \frac{3 }{2 v_2}N^\text{CW}_{\psi_1}\\
    \delta \mathrm{Re}\left(m_{12}^2\right) &=  t_1 v_1 v_2 + H^\text{CW}_{\psi_1, \psi_2}\\
    \delta \lambda_1 & = -\frac{v_2^2}{v_1^2}  t_1 + \frac{1}{v_1^2} H^\text{CW}_{\psi_1, \psi_1} - \frac{1}{v_1^2} H^\text{CW}_{\zeta_1, \zeta_1}\\
    \delta \lambda_2 & = -\frac{ v_1^2}{v_2^2} t_1 + \frac{1}{v_2^2}H^\text{CW}_{\psi_2, \psi_2} - \frac{1}{v_2^2}H^\text{CW}_{\zeta_2, \zeta_2}\\
    \delta \lambda_3 & = -t_1 + \frac{H^\text{CW}_{\eta_1, \eta_2}}{v_1 v_2} - \frac{1}{v_1^2}H^\text{CW}_{\eta_2, \eta_2} + \frac{H^\text{CW}_{\psi_2, \psi_2}}{v_1^2} - \frac{H^\text{CW}_{\zeta_1, \zeta_2}}{v_1 v_2}\\
    \delta \lambda_4 & = t_1 + \frac{2 }{v_1^2}H^\text{CW}_{\eta_2, \eta_2} - \frac{2 }{v_1^2}H^\text{CW}_{\psi_2, \psi_2}\\
    \delta \mathrm{Re}\lambda_5 & = t_1 \\ 
    \delta \mathrm{Im}\lambda_5 & = -\frac{2  v_1}{v_2}~t_2 - \frac{2 }{v_2^2}H^\text{CW}_{\zeta_1, \psi_1}\\
    \delta T_\text{CB} &= -N^\text{CW}_{\rho_2}\\
    \delta T_1 & = v_1 H^\text{CW}_{\psi_1, \psi_1} + v_2 H^\text{CW}_{\psi_1, \psi_2} - N^\text{CW}_{\zeta_1}\\
    \delta T_2 & = v_1 H^\text{CW}_{\eta_1, \eta_2} + v_2 H^\text{CW}_{\eta_2, \eta_2} - N^\text{CW}_{\zeta_2}\\
    \delta T_\text{CP} & = -\frac{v_1 }{v_2} N^\text{CW}_{\psi_1}- N^\text{CW}_{\psi_2}\\
    \delta \mathrm{Im}\lambda_6 & = t_2\\
    \delta \mathrm{Im}\lambda_7 & = \frac{ v_1^2}{v_2^2} t_2+
                                  \frac{v_1
                                  }{v_2^3}H^\text{CW}_{\zeta_1,
                                  \psi_1} + \frac{1}{v_1
                                  v_2}H^\text{CW}_{\zeta_2, \psi_2} \label{eq:glge}\;,
  \end{align}
\end{subequations}
with
\begin{align}
H^{\text{CW}}_{\phi_i,\phi_j} &\equiv
\partial_{\phi_i} \partial_{\phi_j}\left.V_{\text{CW}} (\phi)\right|_{\phi=
  \langle\phi^c\rangle_{T=0}} \\
 N^{\text{CW}}_{\phi_i} &\equiv \partial_{\phi_i} \left.V_{\text{CW}}
   (\phi) \right|_{\phi=\langle\phi^c\rangle_{T=0}} \,.  
\end{align}
For the procedure on the treatment of the infrared divergences for the
Goldstone bosons in the Landau gauge that arrive in the second
derivative of the Coleman Weinberg potential
\cite{Cline:1996mga,Cline:2011mm,Dorsch:2013wja,Camargo-Molina:2016moz,Martin:2014bca,Elias-Miro:2014pca,
  Casas:1994us} we refer to Ref.~\cite{Basler:2016obg}.

\section{Calculation of the Electroweak Baryogenesis}
\label{sec:baryocalc}
Before we go into the details of the computation of the BAU, $\eta$,
we first sketch the general idea of EWBG. The EWPT triggers the
expansion of bubbles that contain the broken phase with a
non-vanishing VEV $\langle \phi \rangle \ne 0$ within the surrounding
symmetric phase  with $\langle \phi \rangle = 0$. CP-violating
interactions generate a net-asymmetry of the left-handed
fermions in front of the bubble wall. Baryon-number violating
sphaleron processes convert the
left-handed fermions into baryons and vice versa. While the
bubbles are expanding the baryons diffuse through the bubble
wall. Inside the bubble, in the broken phase, the sphaleron decay rate
is strongly suppressed so that the conversion between baryons and
left-handed particles does not continue. The suppression requires an
EWPT that is of strong first order. The criterion for a strong
first-order EWPT is given by $\xi_c = v_c/T_c \gtrsim 1$
\cite{Quiros:1994dr,Moore:1998swa}, where $v_c$ 
denotes the critical VEV at the critical temperature $T_c$. The
critical temperature $T_c$ is defined as the temperature where two
degenerate global minima exist. \s

In order to determine $\xi_c$ we compute the loop-corrected
effective potential at finite temperature. Since the effective potential at 
finite temperature was already discussed in
full detail in \cite{C2HDM,Basler:2018cwe,Basler:2019iuu} for the
C2HDM including the presentation of the adapted renormalization scheme and the thermal corrections, we skip the discussion here
and refer to the previous works. Still, we want to make two remarks here.
In contrast to the value of the effective potential at the minimum,
the VEV determined from the effective potential is gauge
dependent. The issue of gauge dependence has been analysed in the
literature
\cite{gauge1,Buchmuller:1994vy,Laine:1994zq,Patel:2011th,Wainwright:2011qy,Wainwright:2012zn,Garny:2012cg,Laine:2017hdk}. Gauge-invariant
approaches have been proposed within simpler models applying certain
approximations. While a gauge-invariant treatment for the analysis of the EWPT would
certainly be preferred, this is beyond the scope of this paper. The
effective potential also depends on the renormalization scale
$\mu$. For discussions of the effective three-dimensional theory instead
of the conventional perturbative approach, we refer to
\cite{Linde:1980ts,Ginsparg:1980ef,Appelquist:1981vg,Farakos:1994kx,Losada:1996ju,Laine:1997dy,Laine:1998qk,Andersen:1998br,Csikor:1998eu,Laine:1999rv,Laine:2000rm,Laine:2012jy,Gorda:2018hvi,Kainulainen:2019kyp}. \s

In the following we present the calculation of the actual
BAU, $\eta$, to set the applied conventions and
notation. In this analysis two non-local approaches for the
determination of $\eta$ are compared. The first approach is based on the
semi-classical force
\cite{Cline1998,Kainulainen2001,Fromme:2006wx,Fromme:2006cm} yielding a set
of fluid equations. We will refer to this approach as
\FH. The \FH ansatz works for
\textit{thick} bubble walls, so that the wall thickness $L_W$ is assumed to be
larger than the typical de-Broglie wavelength of the particles in
front of the bubble wall. The typical wavelength of a particle in the
plasma is given by the inverse temperature $T^{-1}$ implying that the
Wenzel-Kramers-Brillouin (WKB)
approach used in \FH is valid for bubble walls with
\begin{equation}
	1\ll L_W T_c \,,
	\label{ThickCondition}
\end{equation}
where $T_c$ denotes the critical temperature at which the electroweak
phase transition takes place. Additionally, only small wall velocities
are assumed in\cite{Fromme:2006wx,Fromme:2006cm}. This allows us to
simplify the resulting transport equations further. As mentioned
recently in Ref.~\cite{Cline:2020jre} this can be generalized to arbitrary
wall velocities even above the speed of sound of the plasma. The ansatz
for arbitrary wall speeds is left for further future investigations.\s

The second approach is based on the competing VEV-insertion
approximation (\VIA) \cite{Riotto:1995hh, Riotto:1997vy,
  Lee:2004we,Chung:2009qs}. \VIA formulates 
the quantum transport equations in the Closed Time
Path (CTP) or Schwinger-Keldysh formalism \cite{Schwinger:1960qe,Keldysh:1964ud,Chou:1984es}. To extract the
respective source terms the fermionic two-point
functions of the corresponding particles are expanded at leading order (LO) in the
spatially varying Higgs field VEV $v(z)$, where $z$
denotes the perpendicular distance to the wall.
The next-to-leading order (NLO) contributions to
the CP-violating source terms and the relaxation rates
have been calculated recently \cite{Postma:2019scv}, but they are
not used in this analysis.
\VIA also allows us to include additional leptons in the transport
equations such as the $\tau$-lepton \cite{deVries:2018tgs}. Including leptons in
the transport equations has the advantage that
the generated densities
are not suppressed by strong sphaleron interactions and that the chiral
flux of the leptons can diffuse more efficiently in the plasma.
In this way the $\tau$ contributions might enhance the produced BAU.
We will compare different \VIA systems including only top quarks
($t$), top and bottom quarks ($t+b$) and finally top and bottom quarks as well
as $\tau$ leptons ($t+b+\tau$).
\VIA can be understood as an expansion in $v(z)/T$, whereas \FH
corresponds to an expansion in $\cbrak{L_W T}^{-1}$.
Both approaches rely on the bubble wall
dynamics and its profile.
In the analysis we treat the bubble wall velocity as open parameter
and use the \textit{standard} assumption that the nucleating bubble is
treated in the bubble rest frame and approximated
by a planar wall so
that the only parameter needed in both approaches is the wall profile
depending on the space-time coordinate $z$ referring to the wall
distance. Furthermore, we use a \textit{two-step} approach in both
cases to calculate $\eta$.
In the first step we solve the (quantum) transport equations for the
left-handed fermion excess $n_L$ in front of the
bubble wall, and in the 
second step this fermion asymmetry triggers the generation of the
baryon asymmetry via the electroweak sphaleron transition.

\subsection{Calculation of the Bubble Wall Profile}
To describe the bubble wall profile the kink solution is used by which
the VEV profile as a function of the bubble wall
distance $z$ is described as \cite{Fromme:2006cm,Fromme:2006wx}
\begin{equation}
	f(z) = \frac{f_0}{2}\cbrak{1-\tanh\frac{z}{L_W}}\,,
	\label{wall_profile}
\end{equation}
where $f(z)$ is the value of the VEV at given $z$ and $f_0$ the value
of the VEV inside the broken phase. Furthermore, the wall thickness
$L_W$ is given by \cite{Fromme:2006wx}
\begin{equation}
	L_W = \frac{v_c}{\sqrt{8 V_b}}\,,
\end{equation}
with $v_c$ being the critical VEV at the electroweak phase transition
and $V_b$ the barrier height between both degenerate global minima
(at the critical temperature $T_c$). The numerical
values of the critical VEV $v_c$ and the critical temperature
$T_c$ for a given parameter point are obtained from \ewbgBSMPT, which
also calculates $L_W$.
For this, \ewbgBSMPT determines the tunnel path between both global
minima numerically.
Starting with the direct connection between both minima, the straight
path as a first guess for the tunnel path can be parametrised as
\begin{align}
	\quad \vec{\omega}(t) & = \vec{\omega}_s + t \vec{n}       \\
	\vec{n}               & = \vec{\omega}_b-\vec{\omega}_s\,,
\end{align}
where $\vec{\omega}_{s/b}$ is the VEV configuration of the symmetric
and broken minimum, respectively, and $t\in\left[0,1\right]$.
Successively, the global minima in the orthogonal planes along the
straight path are determined. They form a grid that approximates the
tunnel path between the two degenerate minima. 
The barrier height $V_b$ is then obtained as
the difference between the maximum value of the effective potential
along this path and the value of the effective potential at $v_c$. 
\begin{figure}
	\centering
	\includegraphics[width=0.6\textwidth]{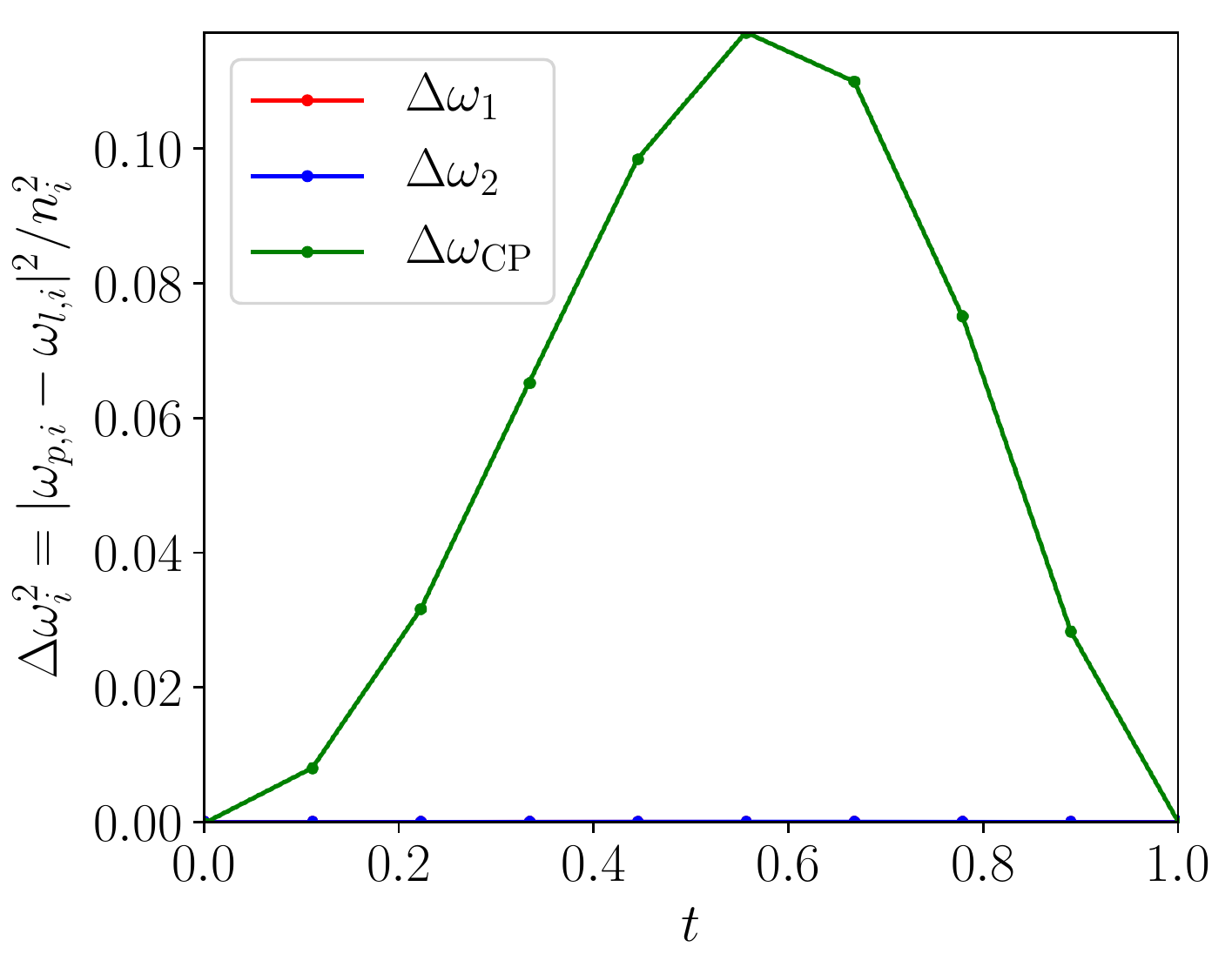}
	\caption{Normalized deviation from the
          straight line path at each step $i$ along the path where
		$\vec{\omega}_l$ are the points along the direct line from
		$\vec{\omega}_s$ to $\vec{\omega}_b$. $\vec{\omega}_p$ refers to
		the found global minimum in the orthogonal planes. The
                color code denotes the various possible minima, $\omega_1$
                (red), $\omega_2$ (blue) and $\omega_{\text{CP}}$
                (green). Red and blue points are on top of each other
                and are almost zero.
	}
	\label{tunnelpath}
\end{figure}
For a more detailed discussion of the numerical method we refer to the
manual of \ewbgBSMPT. In \cref{tunnelpath} the result for one example
parameter point of the C2HDM\footnote{A similar behaviour is observed
	for all parameter points used in the numerical analysis.} is
given. The figure displays the squared difference
between the found VEV vector
$\vec{\omega}_p (t)$ of the tunnel path and the straight connection
$\vec{\omega}_l (t)$ between both minima normalized to the squared
difference between the broken and symmetric VEVs at each step $i$
  (corresponding to discrete values of $t\in [0,1]$),
\begin{equation}
	\Delta \omega_i^2 =
	\frac{|\vec{\omega}_{p,i}-\vec{\omega}_{l,i}|^{2 }}{n_i^2} \;,
\end{equation}
for $\omega_i=\omega_1$ (red), $\omega_2$ (blue) and
$\omega_{\text{CP}}$ (green), versus the parameter $t$.
Both electroweak
VEVs $\omega_1$ and $\omega_2$ do not show any deviation from the
direct connection, only for the CP-violating VEV of the C2HDM,
$\omega_{\text{CP}}$, there is a significant deviation. Similar
observations were made 
in \cite{Fromme:2006cm} showing that the straight line is a good first
approximation of the tunnel path, but the CP-violating VEV was not
taken into account in \cite{Fromme:2006cm}. By determining $L_W$ and
the general VEV configuration at the critical temperature\footnote{To
	be precise one has to take into account that the actual bubble
	formation takes place at the nucleation  temperature $T_N<T_c$. As a
	first approximation we use the critical temperature.} the bubble
wall profile can be parametrised in both approaches, \VIA and \FH, by
using \cref{wall_profile}. \s

Both approaches require the complex phases of
the leptons and quarks as a function of the bubble wall distance $z$.
In the following, we give the explicit formulas
for the C2HDM case by making use of \cref{wall_profile}.
Through the possibility of a CP-violating VEV at
non-zero temperature the quark and lepton masses can become
complex.\footnote{While a CP-violating mass can be avoided by a
	redefinition of the fermion field, this redefinition then only
	applies to the temperature value at which it is performed and not
	for all temperatures under investigation.}
For the type I version of the C2HDM where all leptons and quarks are
coupled to the second Higgs doublet $\Phi_2$, the Yukawa Lagrangian
yields the following mass terms
\begin{equation}
	m_i(z)= \frac{y_{i}}{\sqrt 2} \cbrak{\wtwo(z)-\ii \wCP(z)}\,,
	\label{zmass}
\end{equation}
where $y_i$ is the zero-temperature Yukawa coupling of the respective
particle $i$ and $\wtwo$, $\wCP$ the VEVs defined in
Eq.~(\ref{eq:2hdmdoublets1}). For the type II C2HDM
\cref{zmass} only holds for the up-type quarks. For the leptons and
down-type quarks the VEV $\wone$ of the Higgs doublet $\Phi_1$ gives
rise to the mass term and no complex part is introduced there, so that the masses are
real.
Defining the complex mass of the particle $i$ as
\begin{equation}
	m_i (z) = \frac{y_{i}}{\sqrt 2} \cbrak{\wtwo(z)-\ii \wCP(z)} =
	\frac{y_i}{\sqrt{2}} \sqrt{\wtwo^2+\wCP^2} \exp \cbrak{\ii
		\arg\cbrak{\wtwo - \ii \wCP}}\equiv \vert m_i(z)\vert
	\exp(\ii\theta^{(i)}(z))\,,
	\label{massdef}
\end{equation}
allows us to parametrize the phase evolution as a function of the wall distance as
\begin{equation}
	\theta^{(i)}(z) = \cbrak{\theta_{\text{brk}}^{(i)} - \frac{\theta_{\text{brk}}^{(i)}-\theta_{\text{sym}}^{(i)}}{2} \cbrak{1+\tanh\frac{z}{L_W}} }\,,
\end{equation}
where $\theta_{\text{brk}}$ is the phase in the broken minimum and
$\theta_{\text{sym}}$ the one in the symmetric minimum,
respectively. 
Note that the phase in the totally symmetric minimum with all VEVs
vanishing is arbitrary, so that we chose $\theta_{\text{sym}}$ as the
phase of the symmetric minimum plus an infinitesimal shift along the
tunnel path so that we have 
a smooth phase along the tunnel path. For further details, we refer to
\cite{Basler:2020nrq}. The broken and symmetric phases for the
particle $i$ are given by 
\begin{equation}
	\theta_{\text{brk}}^{(i)} =  \arg\cbrak{\omega_{2,c}^{(i)}-\ii
		\omega_{\text{CP},c}^{(i)}}\,,\quad \theta_{\text{sym}}^{(i)} =
	\arg\cbrak{\omega_{2,\text{s}}^{(i)}-\ii\omega_{\text{CP,s}}^{(i)}}\,,
\end{equation}
where the index $c$ denotes the critical VEVs and $s$ the VEVs in
  the symmetric phase.
For a more detailed description of the numerical approach we refer
again to the manual of \ewbgBSMPT \cite{Basler:2020nrq}. 

\subsection{Semi-classical Force Approach}
The semi-classical force method uses the existence of a complex fermion mass. This
complex mass induces in the presence of a varying Higgs background a
semi-classical force term which can be deduced by applying the WKB
approximation \cite{Fromme:2006cm,Fromme:2006wx,Cline:2020jre}, or
from the closed-time-path (CTP) formalism of thermal field theory
\cite{Kainulainen:2001cn,Kainulainen:2002th,Prokopec:2003pj,Prokopec:2004ic},
yielding
\begin{equation}
	F = - \frac{\cbrak{m^2}^{\prime}}{2E_0} \pm s \frac{\primed{\cbrak{m^2\theta^{\prime}}}}{2 E_0 E_{0,z}}\mp \frac{\theta^{\prime}m^2\cbrak{m^2}^{\prime}}{4 E_0^3 E_{0,z}} \,,
	\label{SK_Force}
\end{equation}
where $E_0 $ is the conserved energy of the quasi-particles
in front of the bubble wall in the rest frame of the wall,
$E_{0,z}^2=E_0^2-\vec{p}_{\parallel}$ with the momentum $\vec
	p_{\parallel}$ parallel to the bubble wall and $\cbrak{\dots}^{\prime}$
denotes the derivative with respect to the wall distance $z$. For
better readability we skipped the $z$ dependences of $m,\theta, E_0$, and $E_{0,z}$
in \cref{SK_Force}.
The mass $m$ and the phase $\theta$ are defined as in \cref{massdef}
and $s$ denotes the spin of the particle.
The first term in \cref{SK_Force} corresponds to the classical
solution since
the particle changes its mass in the varying Higgs background while
moving and conserves CP, whereas the second and third part
besides the spin are dependent
on the particle's nature ($+$ particle/$-$ antiparticle) and therefore
induce CP violation. This part is only present if the particle has a complex
mass phase. Allowing for small kinetic perturbations $\delta f_i$ in
the distribution functions $f_i$ of the particle species $i$ we have
(the $+(-)$ refers to fermions (bosons), $\beta=1/T$),
\begin{equation}
	f_i = \frac{1}{\e^{\beta\left[\gamma_W \cbrak{E_0+v_W
					p_z}-\mu_i\right]}\pm 1} + \delta f_i\,,
\end{equation}
with the Lorentz boost factor $\gamma_W=1/\sqrt{1-v_W^2}$
of the wall. The chemical potential $\mu_i$
describing the departure from chemical equilibrium, allows us 
to express the Boltzmann equations for the near-equilibrium system as
\begin{equation}
	\Lio[f_i]\equiv\cbrak{v_g\partial_z + F \partial_{p_z}}f_i = \mathcal C[f_i]\,,
	\label{Boltzmann_equatio}
\end{equation}
where $\Lio[f_i]$ is the Liouville operator, $v_\mathrm{g}$ the group
  velocity of the WKB wave package given by \cite{Fromme:2006cm}
\begin{align}
	v_\mathrm{g} = \frac{p_z}{E_0} \left( 1 \pm \frac{s}{2}  \primed{\theta}  \frac{m^2}{E_0^2 E_{0,z}} \right) \,, \label{Eq:EWBG:vg}
\end{align}
and $F$ denotes the semi-classical force given in
\cref{SK_Force}. The collision integral $\mathcal C[f_i]$ is model
dependent and can be linked to the interaction rates of the thermal
bath \cite{Cline:2000nw}.
The force term splits into a CP-even and two CP-odd 
terms. Additionally, since the CP-even and CP-odd components are
equal at first order, the perturbations
  $\mu_i$ around the chemical 
equilibrium have to be expanded to the second order in the CP-odd
terms in order to account for CP-violating effects. We therefore
solve the Boltzmann equation separately for
 $\mu_{e/o}$and $\delta f_{i,e/o}$,
\begin{align}
	\mu_i = & \mu_{i,1e} + \mu_{i,2o} + \mu_{i,2e}\,, \qquad \delta f_i = \delta f_{i,1e} + \delta f_{i,2o} + \delta f_{i,2e} \,, \label{Eq:EWBG:ExpansionPot}
\end{align}
where $e(o)$ corresponds to the CP-even (odd) part. The
indices 1 and 2 indicate the order in the gradient expansion used in
\cite{Fromme:2006wx}. To simplify the actual solution of the transport
equation in \cref{Boltzmann_equatio}, only the two
lowest moments of the equation are taken into account, the zero-th and
first moment of \cref{Boltzmann_equatio}. The weighted average are defined as follows for the
zero-th and the first moment, 
respectively,
\begin{equation}
	\braket{X} = \frac{\int d^3p X(p)}{\int d^3
          f_{0+}^{\prime}(m=0)}\,,\quad \braket{\frac{p_z}{E_0}X} =
        \frac{\int d^3p \frac{p_z}{E_0}X(p)}{\int d^3
          f_{0+}^{\prime}(m=0)}\,,
\label{eq:average}
\end{equation}
where the derivative of the massless fermion distribution function
\begin{equation}
	f_{0+}^{\prime}(m=0) \equiv
	f_i|_{i=fermion,\mu_i=0,\delta f_i=0,v_W=0} \label{eq:normfactor}
\end{equation}
is chosen as normalisation\footnote{An
	additional \textit{factorisation assumption} is needed, since the
	momentum dependence of $\delta f$ is not known. In this case one has
	to assume that the average factorises, $\braket{ X \delta f} =
		\left[X \frac{p_z}{E_0}\right]u$, where $u$ is the plasma velocity and
	$\left[\dots\right]$ is the momentum average with the massive
	distribution function.}. 
By defining the plasma velocity
\begin{equation}
	u_i \equiv \braket{\frac{p_z}{E_0}\delta f_i}
\end{equation}
the Liouville operator in \cref{Boltzmann_equatio} on the one side
produces source terms 
and on the other side relates the chemical potentials and plasma velocities with thermal
transport coefficients, denoted $K_i$. 
The only missing piece are
the zero-th and the first moments of the collision integrals
\begin{equation}
	\braket{\mathcal C[f_i]}\quad\text{and}\quad \braket{\frac{p_z}{E_0}\mathcal C[f_i]}\,,
\end{equation}
which can be expressed in terms of the
inelastic and total interaction rates, $\Gamma_{\text{inel}}$ and
$\Gamma_{\text{tot}}$, respectively\cite{Cline:2000nw}, 
\begin{equation}
	\braket{\mathcal C[f_i]} = \Gamma_{\text{inel}} \sum \mu_i\quad\text{and}\quad \braket{\frac{p_z}{E_0}\mathcal C[f_i]} = - \Gamma_{\text{tot}} u\,.
\end{equation}
The second-order CP-odd chemical potential is
given by the difference of the chemical potential of the particle 
and the one of the anti-particle, 
\begin{align}
	\mu_{i,2} ={} & \mu_{i,2o} - \bar{\mu}_{i,2o} \,. \label{Eq:EWBG:ChemDiff}
\end{align}
The index $i$ denotes the involved particle species, given by the top
quark and its charged conjugated, $t$ and $t^c$, the bottom quark $b$
and the Higgs boson $h$.  The
chemical potential of the corresponding 
antiparticle is denoted by $\bar{\mu}_i$.
 The transport equations include Yukawa interactions, strong
sphaleron transitions and $W$-boson scattering.
The top transport equations can then be written as
\cite{Fromme:2006cm,Fromme:2006wx}
\begin{subequations}
	\begin{align}
		0 =   & 3 \vw K_{1,t} \left( \partial_z \mu_{t,2} \right) + 3\vw K_{2,t} \left( \partial_z m_t^2 \right) \mu_{t,2} + 3 \left( \partial_z u_{t,2} \right) \notag
		\\ &- 3\Gamma_y \left(\mu_{t,2} + \mu_{t^c,2} + \mu_{h,2} \right) - 6\Gamma_M \left( \mu_{t,2} + \mu_{t^c,2} \right) - 3\Gamma_W \left( \mu_{t,2} - \mu_{b,2} \right) \notag
		\\ &- 3\Gamma_{ss} \left[ \left(1+9 K_{1,t} \right) \mu_{t,2} + \left(1+9 K_{1,b} \right) \mu_{b,2} + \left(1-9 K_{1,t} \right) \mu_{t^c,2} \right] \label{Eq:TransportEquations:mut} \,,\\
		0 =   & 3\vw K_{1,b} \left(\partial_z \mu_{b,2}\right) + 3 \left(\partial_z u_{b,2} \right) - 3\Gamma_y \left( \mu_{b,2} + \mu_{t^c,2} + \mu_{h,2} \right) - 3\Gamma_W \left( \mu_{b,2} - \mu_{t,2} \right) \notag \label{Eq:TransportEquations:mub}    \\
		      & - 3\Gamma_{ss} \left[ \left( 1 + 9K_{1,t}\right) \mu_{t,2} + (1+9K_{1,b}) \mu_{b,2} + (1-9K_{1,t}) \mu_{t^c,2} \right] \,,                                                                                                                      \\
		0=    & 3 \vw K_{1,t} \left( \partial_z \mu_{t^c,2} \right)  + 3\vw K_{2,t} \left( \partial_z m_t^2 \right)  \mu_{t^c,2} + 3 \left( \partial_z u_{t^c,2} \right) \notag                                                                                 \\
		      & - 3\Gamma_y \left(\mu_{t,2} + \mu_{b,2} + 2\mu_{t^c,2} + 2\mu_{h,2} \right) - 6\Gamma_M \left( \mu_{t,2} + \mu_{t^c,2} \right) \notag                                                                                                           \\
		      & - 3\Gamma_{ss} \left[ \left( 1+9 K_{1,t}\right) \mu_{t,2} + \left(1+9K_{1,b}\right) \mu_{b,2} + \left(1-9K_{1,t}\right) \mu_{t^c,2} \right] \label{Eq:TransportEquations:mutc} \,,                                                              \\
		0 =   & 4\vw K_{1,h} \left( \partial_z \mu_{h,2}\right) +
		4\left( \partial_z u_{h,2}\right) - 3\Gamma_y \left(
		\mu_{t,2} + \mu_{b,2} + 2\mu_{t^c,2} + 2\mu_{h,2} \right) -
		4\Gamma_h
		\mu_{h,2} \label{Eq:TransportEquations:muh} \,,\\
		S_t = & -3K_{4,t} \left( \partial_z \mu_{t,2}\right) + 3\vw \tilde{K}_{5,t} \left( \partial_z u_{t,2}\right) + 3\vw \tilde{K}_{6,t} \left( \partial_z m_t^2 \right) u_{t,2} + 3\Gamma_t^\mathrm{tot} u_{t,2} \label{Eq:TransportEquations:ut} \,,       \\
		0 =   & -3K_{4,b} \left( \partial_z \mu_{b,2} \right) + 3\vw \tilde{K}_{5,b} \left(\partial_z u_{b,2}\right) + 3\Gamma_b^\mathrm{tot} u_{b,2} \label{Eq:TransportEquations:ub} \,,                                                                      \\
		S_t = & -3K_{4,t} \left( \partial_z \mu_{t^c,2}\right) + 3\vw \tilde{K}_{5,t} \left( \partial u_{t^c,2}\right) + 3\vw \tilde{K}_{6,t} \left( \partial_z m_t^2\right) u_{t^c,2} + 3\Gamma_t^\mathrm{tot} u_{t^c,2} \label{Eq:TransportEquations:utc} \,, \\
		0 =   & -4K_{4,h} \left( \partial_z \mu_{h,2} \right) + 4\vw \tilde{K}_{5,h} \left( \partial_z u_{h,2} \right) + 4\Gamma_h^\mathrm{tot} u_{h,2} \label{Eq:TransportEquations:uh} \,,
	\end{align}
	\label{Eq:TransportEquations}
\end{subequations}
with the source term of the top
quark\footnote{Because of the smallness of the bottom quark mass
the source term of the bottom quark can be neglected \cite{Fromme:2006wx}.}
\begin{align}
	S_t = & -v_W K_{8,t} \partial_z \left( m_t^2 \partial_z \theta \right) + v_W K_{9,t} \left( \partial_z \theta \right) m_t^2  \left( \partial_z m_t^2\right) \label{Eq:TransportEquations:Source} \,.
\end{align}
        Analogous to the chemical potential
	Eq.~(\ref{Eq:EWBG:ChemDiff}) the transport equations only depend on
	the differences between the CP-odd components of the plasma
	velocities of the particles, $u_{i,2o}$, and of their antiparticles,
	$\bar{u}_{i,2o}$,
\begin{equation}
	u_{i,2} = u_{i,2o} - \bar{u}_{i,2o} \;.
\end{equation}
For the strong sphaleron rate $\Gamma_{ss}$ and the top Yukawa
  rate $\Gamma_y$ we use\footnote{Note, that for simplicity we use the
  critical temperate and not the nucleation temperature.} \cite{Fromme:2006cm,Huet:1995sh,Moore:1997im} 
\begin{equation}
\Gamma_{ss} = 4.9 \times 10^{-4} T_c \;, \qquad 
\Gamma_y = 4.2 \times 10^{-4} T_c
\end{equation}
The $W$-exchange rate is approximated by the total Higgs interaction rate
$\Gamma_h^{\text{tot}}$ \cite{Fromme:2006cm}.
The Higgs number violating rate is given by \cite{Huet:1995sh}
\begin{equation}
	\Gamma_h = \frac{m_W^2(z,T_c)}{50 T_c}\,,
\end{equation}
where the $W$-mass is determined numerically at given temperature
$T_c$ and wall distance $z$ by {\ewbgBSMPT}. The spin-helicity
flipping rate $\Gamma_M$ for the top quark is implemented as
\cite{Huet:1995sh} 
\begin{equation}
	\Gamma_M = \frac{m_t^2(z,T_c)}{63 T_c}\,,
\end{equation}
where again the mass of the top quark is determined numerically at
given distance and temperature $T_c$. 
The total interaction rates in \cref{Eq:TransportEquations} can be
related to the diffusion constants $D_i$ of the
quarks and Higgs bosons as \cite{Fromme:2006wx,Fromme:2006cm}
\begin{align}
	D_t = \frac{K_{4,t}}{K_{1,t} \Gamma_t^\mathrm{tot}} \,,\quad
	D_b = \frac{K_{4,b}}{K_{1,b} \Gamma_b^\mathrm{tot}}  \,,\quad
	D_h = \frac{K_{4,h}}{K_{1,t} \Gamma_h^\mathrm{tot}}  \,.
\end{align}
The thermal transport coefficients are defined as
\begin{subequations}
	\begin{align}
		K_{1,i} =         & - \Braket{\frac{p_z^2}{E_0} \partial_E^2 f_{i,0} } \,,                                                                  \\
		K_{2,i} =         & \Braket{\frac{\partial_E^2 f_{i,0}}{2E_0} } \,,                                                                         \\
		K_{4,i} =         & \Braket{\frac{p_z^2}{E_0^2} \partial_E f_{i,0} } \,,                                                                    \\
		\tilde{K}_{5,i} = & \left[ \frac{p_z^2}{E_0} \partial_E f_{i,0} \right] \,,                                                                 \\
		\tilde{K}_{6,i} = & \left[ \frac{E_0^2-p_z^2}{2E_0^3} \partial_E f_{i,0} \right] \,,                                                        \\
		K_{8,i} =         & \Braket{\frac{\vert p_z\vert \partial_E f_{i,0}}{2E_0^2 E_{0z}} } \,,                                                   \\
		K_{9,i} =         & \Braket{\frac{\vert p_z\vert}{4E_0^3 E_{0z}} \left( \frac{\partial_E f_{i,0}}{E_0} - \partial_E^2 f_{i,0} \right) } \,,
	\end{align}
	\label{Eq:Kfactors}
\end{subequations}
with the expectation values given by 
\begin{align}
	\Braket{X} = \frac{\int \mathrm{d}^3p X(p)}{\int \mathrm{d}^3p \partial_E f_{0+}(m=0)}  \,,\quad
	\left[X\right] = \frac{\int \mathrm{d}^3p X(p)}{\int \mathrm{d}^3p f_{i,0,\vw}} = \frac{\int \mathrm{d}^3p X(p)}{\int \mathrm{d}^3p f_{i,0}|_{\vw=0}}
\end{align}
and the distribution functions 
\begin{align}
	f_{i,0} = \left. f_{i}\right|_{\mu_i = 0, \delta f_i = 0, \vw = 0} \,,\quad
	f_{0+} = \left. f_{i}\right|_{i=\mathrm{fermion},\mu_i=0,\delta f_i = 0, \vw = 0}\,, \quad
	f_{i,0,v_W} =  f_{i,0} + \vw p_z \partial_{E_0}  f_{i,0}\,.
\end{align}
The first two equations describe the distribution
function in chemical equilibrium. The third one is the Taylor series
of the distribution in chemical equilibrium for small wall velocities.
Note that the assumption of small wall velocities is explicitly used
to simplify the thermal transport coefficients in
\cref{Eq:Kfactors}. To keep the full wall velocity dependence in the
transport equations it is required to adapt the thermal transport
coefficients in \cref{Eq:Kfactors}. The implementation of the full
dependence as discussed in Ref.~\cite{Cline:2020jre} is left for future
work.  \s

In the numerical implementation, the chemical potentials and the
plasma velocities are assumed to vanish at $z\approx 4 L_W$ as a
boundary condition. The factor 4 of the wall thickness has no
physical interpretation and it was checked that this choice in the
numerical set-up has no impact on the end result. Furthermore, with the choice of the
kink profile the corresponding VEV profile already has negligibly
small values at $z\approx 4 L_W$ implying that the choice of this
boundary condition is justified. \s

The differential system of equations in \cref{Eq:TransportEquations}
is solved numerically in \ewbgBSMPT. For that the thermal coefficients
in \cref{Eq:Kfactors} are evaluated numerically for a given
two-dimensional grid in squared mass and temperature and interpolated
as a bi-cubic spline to optimize the run time of the calculation. The
numerical solution of the transport equation system in
\cref{Eq:TransportEquations} is performed by using the numerical {\tt
  c++} library {\tt boost} \cite{BoostLibrary} implemented in {\tt
  \ewbgBSMPT}. Solving the transport equations of
\cref{Eq:TransportEquations} yields the chemical potentials $\mu_i$ of
each particle species. Assuming local baryon number conservation the
chemical potential of the left-handed quarks is then given by
\cite{Fromme:2006cm} 
\begin{align}
	\mu_{B_L} = & \frac{1}{2} \left(1+4K_{1,t}\right) \mu_{t,2} + \frac{1}{2} \left(1+4K_{1,b}\right) \mu_{b,2} - 2K_{1,t} \mu_{t^c,2} \,,
\end{align}
which triggers the
generation of the baryon asymmetry in the
electroweak sphaleron transition. \s

The actual value of the BAU can then be calculated from the solution
of the transport equations since we assume that the weak sphaleron transition rate $\Gamma_{ws}$ is much smaller than all contributing interaction rates of the thermal plasma. In this way we can first calculate the produced left-handed fermion asymmetry in front of the bubble wall and in the second step we calculate the produced baryon asymmetry due to the electroweak sphaleron transition.
The produced BAU can be calculated with \cite{Fromme:2006cm,Fromme:2006wx}
\begin{equation}
	\eta_B = \frac{n_B}{s}=\frac{405 \Gamma_{ws}}{4\pi^2 v_w g_* T_c}\int_0^\infty dz \mu_{B_L}(z) \exp\cbrak{-\frac{45\Gamma_{ws}}{4v_W}}\,,
\end{equation}
with the bubble wall velocity $v_W$, the effective degrees of freedom
of the universe at electroweak temperatures $g_*\simeq 106.75$. For
simplicity we calculate the produced BAU at the critical temperature
$T_c$ and use $\Gamma_{ws}\simeq 1.0\cdot 10^{-6} T_c$. The implementation of the nucleation temperature in
  \texttt{BSMPT}, that should actually be used, is left for future
  work. 
For a detailed theoretical derivation of the top transport equations
in \cref{Eq:TransportEquations} 
we refer to Refs.~\cite{Fromme:2006wx,Fromme:2006cm} and for a detailed
description of the implementation of the top transport equations in
\ewbgBSMPT we refer to the manual of \ewbgBSMPT \cite{Basler:2020nrq}.

\subsection{The VEV-Insertion Approximation}
The VEV-insertion approximation (\VIA) can be understood as an expansion in
$v(z)/T$ in which the fermionic two-point function is expanded in terms of the
VEV. The mass fluctuations induced by the varying Higgs background
($v(z)$) are treated as perturbations that interact with the
thermal bath. By including these thermal interactions CP-conserving
and CP-violating source terms for the right-/left-handed fermion
densities can be found. These sources generate a net-asymmetry between
left- and right-handed fermions in front of the bubble wall, which
again is then translated in the two-step approach into the baryon
asymmetry via the electroweak sphaleron transition. Starting with the
quantum transport equations derived in the finite temperature CTP
formalism
\cite{Schwinger:1960qe,Mahanthappa:1962ex,Bakshi:1962dv,Keldysh:1964ud,Chou:1984es} 
the Schwinger-Dyson equation for a Weyl fermion current can be
derived as \cite{deVries:2017ncy,Lee:2004we}
\begin{align}
	\label{Schwinger_Dyson}
	\partial_{\mu}j_i^{\mu} = - \int d^3z \int\limits_{-\infty}^{\infty} dz_0 \Tr\left[  \Sigma_i^{>}(x,z)S_i^{<}(z,x)- S_i^>(x,z)\Sigma_i^<(z,x)\right. \\\nonumber
	\left. + S_i^<(x,z)\Sigma_i^>(z,x) - \Sigma_i^<(x,z)S_i^>(z,x) \right]\,,
\end{align}
with $i=L,R$ for the left- and right-handed fermion, respectively. The
Wightman functions $S^{\lambda}$ ($\lambda=>,<$) and
the corresponding self-energies
$\Sigma^{\lambda}$ are defined in \cite{Lee:2004we}. Again the bubble
is assumed to be planar and the reference frame is the bubble rest
frame which allows us to reformulate the left-hand side of
\cref{Schwinger_Dyson}. By using the diffusion approximation and
Fick's law the left-/right-handed current of
the particle species $i$ can be expressed in terms of the distribution
function $n_{L/R,i}$ of the left-/right-handed particle species $i$,  
\begin{equation}
	\partial_{\mu}j^{\mu}_{R/L,i}(x)\approx v_w \primed{n_{R/L,i}}- D_{R/L,i} \nabla^2 n_{R/L,i} \approx v_w\primed{n_{R/L,i}}-D_{R/L,i} n_{R/L,i}^{\prime\prime}\,,
	\label{Fick_Law}
\end{equation}
with the diffusion constant $D_{R/L,i}$ for the respective particle $i$ and
$\primed{\cbrak{\dots}}$ corresponding to the derivative with respect
to the wall distance $z$. 
The thermal corrections and the complex phases of the masses allow us to write
the mass terms of the quarks/leptons $\Psi$ as follows
\begin{equation}
	\mathcal L \supset - \frac{f_i\cbrak{T,\phi_b}}{\sqrt 2} \bar{\Psi}_L \Psi_R +\hc\,,
	\label{param}
\end{equation}
where $f_i \cbrak{T,\phi_b}\in\mathbb C$ parametrizes the interaction
strength as a function of the Higgs background field $\phi_b$ and
the temperature $T$. Using \cref{param} allows us to cast the right-hand side
of \cref{Schwinger_Dyson} in a CP-conserving part and a CP-violating
part \cite{Lee:2004we}
\begin{equation}
	\text{RHS of \cref{Schwinger_Dyson}}  = S_{CP}^{(i)} + S_{\cancel{CP}}^{(i)}\,,
\end{equation}
with the right-handed CP-violating source term for the particle species $i$ given by
\begin{align}
	\label{CPv_source}
	S_{\cancel{CP}}^{(i)} = \frac{N_c v_w}{\pi^2}\Im\cbrak{\primed{f_i}f_i^*} \int \frac{k^2dk}{\omega_L\omega_R}\Im\left[\frac{\cbrak{n(\epsilon_L)-n(\epsilon_L^*)}}{\cbrak{\epsilon_L-\epsilon_L^*}^2}\cbrak{\epsilon_L\epsilon_R^*-k^2}\right. \\\nonumber
	\left.+\frac{\cbrak{n(\epsilon_L)+n(\epsilon_R)-1}}{\cbrak{\epsilon_L+\epsilon_R}^2}\cbrak{\epsilon_L\epsilon_R+k^2}\right]\,,
\end{align}
with the color factor $N_c=3(1)$ for quarks (leptons) and the
4-momentum $k$ of the fermions. $\primed{f_i}$ corresponds to the
derivative of the interaction strength in \cref{param} with respect to
the wall distance $z$. The left-/right-handed dispersion relation
reads\footnote{Note that for better readability we have
    dropped the index $i$ in the quantities of the integral.}
\begin{equation}
	\epsilon_{L/R}^{i} =\omega_{L/R}^i + \ii \Gamma_{T,L/R}^i \equiv \sqrt{k^2 +\cbrak{m^i_{T,L/R}}^2}-\ii \Gamma^i_{T,L/R}\,\quad \,,
\end{equation}
with the thermal mass $m^i_{T}~\cbrak{i=t,b,\tau}$, the thermal decay width
$\Gamma_{T}^i$ and the Fermi-Dirac distribution
$n(x)=\cbrak{\e^x+1}^{-1}$. For simplicity we assume
the thermal widths to be approximately degenerate for left- and
  right-handed particles,
\begin{equation}
	\Gamma^{i}_{L,T}\approx\Gamma^{i}_{T,R}\approx\Gamma^{i}_{T}\approx 0.16 T\,,
\end{equation}
and for the thermal masses we use \cite{Postma:2019scv}
\begin{align}
	  & \cbrak{m_{T,R}^q}^2 = \cbrak{\frac{g_1^2}{18} + \frac{g_3^2}{6} + \frac{y_q^2}{8}}T^2\,,                                                     \\
	  & \cbrak{\delta m^q}^2 \equiv \cbrak{m_{T,R}^q}^2-\cbrak{m_{T,L}^q}^2 =\cbrak{\frac{5 g_1^2}{96} - \frac{3g_2^2}{32} + \frac{y_q^2}{16}}T^2\,, \\
	  & \cbrak{m_{T,R}^l}^2 = \cbrak{\frac{g_1^2}{8} + \frac{y_q^2}{8}}T^2\,,                                                                        \\
	  & \cbrak{\delta m^l}^2 \equiv \cbrak{m_{T,R}^l}^2-\cbrak{m_{T,L}^l}^2 =\cbrak{\frac{3 g_1^2}{32} - \frac{3g_2^2}{32} - \frac{y_q^2}{16}}T^2\,,
\end{align}
where $q$ corresponds to the quark and $l$ to the lepton type,
respectively. The gauge couplings $g_i$ ($i=1,2,3$) are those of the SM gauge groups
$\mathrm{SU}(3)\times\mathrm{SU_L}(2)\times U(1)_Y$ and $y_q$ denotes
the Yukawa coupling of the respective quark.
For quarks, the difference of the left- and right-handed thermal
masses is not significant and taking the limit of exactly degenerate masses in
\cref{CPv_source} would be a valid approximation. For leptons,
however, this is not the case. Therefore, we expand
\cref{CPv_source} for small $\cbrak{\delta m^i}^2$ of quarks and
leptons. As a second step we exploit $\Gamma_T \ll T$ allowing us to
simplify \cref{CPv_source} significantly, leading to
\begin{align}
	S_{\cancel{CP}}^{(i)} = \frac{N_c v_w}{\pi^2}\Im\cbrak{\primed{f_i}f_i^*}\int dk\frac{k^4}{\omega^4} \left[-\frac{\Gamma_T}{2 \omega}+\frac{5 \Gamma_T}{4 \omega^2}\delta\omega + \cbrak{\frac{\Gamma_T}{\omega}-\frac{5\Gamma_T \delta\omega}{2 \omega^2} }n(\omega)\right. \\\nonumber
	+\left.\cbrak{\frac{-\omega^2}{2\Gamma_T}+\frac{\omega^4}{2k^2\Gamma_T}-\frac{\Gamma_T}{2}+\cbrak{\frac{\omega}{2\Gamma_T}+\frac{3\Gamma_T}{2\omega}}\delta\omega}\primed{n}(\omega)\right]                                                                                  \\\nonumber
	+\mathcal O\cbrak{\delta\omega^2;\cbrak{\frac{\Gamma_T}{T}}^2;n^{\prime\prime}}\,,
\end{align}
with the shorthand notation
\begin{equation}
	\delta\omega = \frac{\cbrak{\delta m}^2}{2\sqrt{k^2+\cbrak{m_R}^2}}\,.
\end{equation}
Note that, for better readability, we again neglect the index $i$.
The CP-conserving interactions $S_{CP}$ in \cref{Schwinger_Dyson}
contain the Yukawa interaction rates,
where we use the approximation 
of\cite{Joyce:1994zn}, 
\begin{equation}
	\Gamma_y^{\text{quark}} \approx 0.19 \alpha_s y_q^2
	T\,\quad\text{,}\,\quad \Gamma_y^{\text{lepton}}\approx
	0.28\alpha_w y_{\tau}^2T\,,
\end{equation}
with the zero-temperature Yukawa couplings $y_q$ and $y_{\tau}$ of the
quarks and $\tau$ leptons, respectively. $\alpha_s$ corresponds to the
strong coupling and $a_w$ to the $SU(2)$ coupling of the SM. The CP-conserving source term reads 
\begin{equation}
	S_{CP} = \Gamma^+_M+ \mu_+ + \Gamma^-_M \mu_-\,, 
\end{equation}
with $\mu_\pm = \mu_L\pm \mu_R$ and  the relaxation rates are given by\cite{Lee:2004we}
\begin{align}
	\label{relax_rate}
	\Gamma_M^{\pm,(i)} = \frac{6}{T^2}\cdot \frac{N_c}{2\pi^2T}\vert f_i\vert^2 \int \frac{k^2dk}{\omega_L\omega_R}\Im\left[-\frac{\cbrak{h(\epsilon_L)\mp h(\epsilon_R^*)}}{\epsilon_R^*-\epsilon_L}\cbrak{\epsilon_L\epsilon_R^*-k^2}\right. \\\nonumber
	\left.+\frac{\cbrak{h(\epsilon_L)\mp h(\epsilon_R)}}{\epsilon_L+\epsilon_R}\cbrak{\epsilon_L\epsilon_R+k^2}\right]\,,
\end{align}
where $h$ denotes the derivative of the Fermi-Dirac distribution
given by
\begin{equation}
	h(x) = \frac{\e^{x}}{\cbrak{\e^x +1}^2}\,.
\end{equation}
Note that the relaxation rate in \cref{relax_rate} is actually rescaled
due to the high temperature expansion of the chemical,
potential 
\begin{equation}
	n = \frac{T^2}{6}\mu \kappa + \mathcal O(\mu_i^3)\,,
	\label{high_temp_chemical_potential}
\end{equation}
that enters the transport equations, with $n$ denoting the number
  density and $\kappa$
the statistical factor for fermions (F,$+$) and bosons
(B,$-$), respectively\footnote{The numerical values for the
  normalisation are $c_F = 6$ and $c_B=3$.}\cite{Chung:2009qs},
given by
\begin{equation}
	\kappa(x) = \kappa_i(0) \frac{c_{F,B}}{\pi^2}
        \int\limits_{m/T}^{\infty} dx \frac{x \e ^x}{\cbrak{\e^x\pm
            1}^2} \sqrt{x^2-m^/T^2}\,. \label{eq:stat}
\end{equation}
As before, we apply the expansion in small mass differences
$\cbrak{\delta m}^2$ and for $\Gamma_T\ll T$ in \cref{relax_rate}, simplifying the 
integration significantly,
\begin{align}
	\Gamma^{-,(i)}_M = \frac{6}{T^2}\cdot \frac{N_c}{2\pi^2T}\vert f_i\vert^2 \int \frac{ dk k^2  }{\omega^2} \cbrak{-\frac{k^2}{\Gamma_T}+\frac{\omega^2}{\Gamma_T} + \frac{k^2\Gamma_T}{\omega^2} + \cbrak{\frac{k^2}{\omega\Gamma_T}-\frac{2 k^2 \Gamma_T}{\omega^3}}\delta\omega} h_f(\omega) \\\nonumber
	+\mathcal O\cbrak{\delta\omega^2;\cbrak{\frac{\Gamma_T}{T}}^2;h_f^{\prime}}\,\,.
\end{align}
Note that
\begin{equation}
	\Gamma^{+,(i)}\sim \cbrak{\dots}\delta\omega \cdot
	\primed{h_f}(\omega)+\mathcal
	O\cbrak{\delta\omega^2;\cbrak{\frac{\Gamma_T}{T}}^2;h_f^{\prime}} \;.
\end{equation}
We drop $\Gamma^{+,(i)}$ for simplicity and also to be consistent with
Ref.~\cite{deVries:2018tgs} which we follow for the
formulation of the full set of transport
equations \cite{deVries:2018tgs}.
The net number density, {\it i.e.}~the number density of
particles minus antiparticles, is denoted as follows
\begin{subequations}
	\begin{align}
		  & n_q = n_{t_L} + n_{b_L}\,,      &   & n_t = n_{t_R}\,,         &   & n_b = n_{b_R}\,,    \\
		  & n_l = n_{\nu_L} + n_{\tau_L}\,, &   & n_{\tau} = n_{\tau_R}\,, &   & n_{\nu} = n_{\nu_L} \,,\\
		&n_{h_k} = n_{h^0_k}+n_{h^{\pm}_k}\,,
	\end{align}
\end{subequations}
where $n_{X_{L/R}}$ is the distribution function of the left- or
right-handed particle species $X$. The index $k$ denotes the doublets $\phi_k = (h_k^{\pm},h_k^0)$. The strong sphaleron rate allows us to relate the densities of the light quarks via
\begin{equation}
	n_{q_1} = n_{q_2} = -2 n_u = - 2 n_d = -2 n_s = -2n_c\,,
\end{equation}
so that only one of them needs to be considered, which we choose to be $n_u$.
Note that the distribution
functions are used and not the chemical potentials. The question which
interactions should be included in the transport equations depends on
the time scale of the diffusion process. By assuming the
\textit{two-step} approach, first the generated left-handed asymmetry
in front of the bubble wall is calculated and in the second step this
asymmetry is translated to the actual baryon asymmetry via an
electroweak sphaleron transition.
The diffusion system is then given by \cite{deVries:2018tgs}
\begin{subequations}
	\begin{align}
		  & \partial_{\mu}j_q^{\mu} = + \Gamma_M^{(t)} \mu_M^{(t)} + \Gamma_M^{(b)} \mu_M^{(b)} + \Gamma_y^{(t)} \mu_Y^{(t)}+ \Gamma_y^{(b)} \mu_Y^{(b)} - 2 \Gamma_{ss} \mu_{ss} - S_{\cancel{CP}}^{(t)} - S_{\cancel{CP}}^{(b)}\,, \\
		  & \partial_{\mu}j_t^{\mu} = - \Gamma_M^{(t)}\mu_M^{(t)} - \Gamma_y^{(t)} \mu_Y^{(t)}+ \Gamma_{ss}\mu_{ss} + S_{\cancel{CP}}^{(t)}\,,                                                                                       \\
		  & \partial_{\mu}j_b^{\mu} = - \Gamma_M^{(b)}\mu_M^{(b)} - \Gamma_y^{(b)} \mu_Y^{(b)}+ \Gamma_{ss}\mu_{ss} + S_{\cancel{CP}}^{(b)}\,,                                                                                       \\
		  & \partial_{\mu}j_l^{\mu} = + \Gamma_M^{(\tau)}\mu_M^{(\tau)} + \Gamma_y^{(\tau)}\mu_Y^{(\tau)} - S^{(\tau)}_{\cancel{CP}}\,,                                                                                              \\
		  & \partial_{\mu}j_{\nu}^{\mu} = 0\,,                                                                                                                                                                                       \\
		  & \partial_{\mu}j_{\nu}^{\mu} = -\Gamma_M^{(\tau)}\mu_M^{(\tau)} - \Gamma_y^{(\tau)}\mu_Y^{(\tau)} + S_{\cancel{CP}}^{(\tau)}\,,                                                                                           \\
		  & \partial_{\mu}j_{h_k}^{\mu} = + \Gamma_y^{(t)} \mu_Y^{(t)} -\Gamma_y^{(b)}\mu_Y^{(b)} + \Gamma_y^{(u)}\mu_Y^{(u)}-\Gamma_y^{(\tau)}\mu_Y^{(\tau)}\,,                                                                     \\
		  & \partial_{\mu}j_u^{\mu} = + \Gamma_{ss} \mu_{ss}\,,
	\end{align}
	\label{VIA_Transport}
\end{subequations}
with the Yukawa rates $\Gamma_y^{(i)}~\cbrak{i=t,b,\tau}$, the relaxation rates
$\Gamma_M^{(i)}$ (defined as $\Gamma^{-,(i)}_M$ in \cref{relax_rate}), the
strong sphaleron rate
\begin{equation}
	\Gamma_{ss} = 14 \alpha_s^4 T_c\,,
\end{equation} and the respective source terms
$S^{(i)}_{\cancel{CP}}$. Note that the light leptons decouple completely
from the system, since there is no corresponding strong sphaleron
interaction. 
It is also possible to neglect the $\tau$ lepton in the system of
transport equations by setting the associated Yukawa rates to
zero. Analogously, the bottom quark can be decoupled.
Assuming $m_b\approx 0$ one can then also drop $u$ in the system
of transport equations 
due to the relation $u=b$ for massless bottom quarks.
The rescaled chemical potentials in \cref{VIA_Transport} are given by
\begin{subequations}
	\begin{align}
		  & \mu_M^{(t)} =\cbrak{\frac{n_t}{\kappa_t} - \frac{n_q}{\kappa_{q}}}\,,            &   & \mu_Y^{(t)} = \cbrak{\frac{n_t}{\kappa_t}-\frac{n_q}{\kappa_q}-\sum_k\frac{h_k}{\kappa_{h_k}}}\,,                \\
		  & \mu_M^{(b)}= \cbrak{\frac{n_b}{\kappa_b} - \frac{n_q}{\kappa_{q}}}\,,            &   & \mu_Y^{(b)} = \cbrak{\frac{n_b}{\kappa_b}-\frac{n_q}{\kappa_q}+\sum_k\frac{h_k}{\kappa_{h_k}}}\,,                \\
		  & \mu_M^{(\tau)}= \cbrak{\frac{n_{\tau}}{\kappa_\tau} - \frac{n_l}{\kappa_{l}}}\,, &   & \mu_Y^{(\tau)} = \cbrak{\frac{n_{\tau}}{\kappa_\tau}-\frac{n_l}{\kappa_l}+\sum_k\frac{n_{h_k}}{\kappa_{h_k}}}\,, \\
		&\mu_{ss} = \cbrak{\frac{2 n_q}{\kappa_q} -\frac{n_t}{\kappa_t} - \frac{n_b}{\kappa_b}-\frac{8 n_u}{\kappa_{L}}-\frac{4 n_u}{\kappa_R}}\,,
	\end{align}
	\label{eq::rescaledMU}
\end{subequations}
where the statistical factor $\kappa_i$ is defined in 
but: Eq.~(\ref{eq:stat}). Note that $\kappa_{L/R}$ refers to left-/right-handed massless quarks, respectively.
Using \cref{Fick_Law} allows us to express the system of transport
equations in \cref{VIA_Transport} as a system of second
order coupled differential equations (ODE). The ODE is solved
by using the {\tt C++} library {\tt Boost::Odeint}\cite{Boost1.66} which we embedded in the {\tt BSMPT} framework. For
technical details of the numerical solution of the ODE we refer to
\cite{Basler:2020nrq}.

The solution of the quantum transport equations given in
\cref{VIA_Transport} allows us to calculate the produced BAU in a
second step. In this second step, the thermalization of the
left-handed excess in front of the bubble to baryons through sphaleron
transitions has to be solved. The thermal system is described by the
differential equation \cite{deVries:2018tgs} 
	\begin{equation}
		-v_W \primed{n_B} - D n_B^{\prime\prime} =- N_f \Gamma_{ws}\cbrak{\mu_{ws}+\mu^0_{ws}}\,,
	\end{equation}
with the bubble wall velocity $v_W$, the baryon asymmetry
distribution function $n_B$, the diffusion constant $D$, the family
number $N_f$ and the electroweak sphaleron transition rate
$\Gamma_{ws}$. The chemical potentials in the last bracket split into
two parts. The first part describes the chemical potentials of the left-handed fermions which are
dynamically produced during the thermalization. The second part
$\mu_{ws}^0$ denotes the initial condition, which biases the
electroweak sphaleron transitions in the first place. Hence, the
initial condition $\mu_{ws}^0$ is given by the sum of all left-handed
fermionic chemical potentials of the solution of \cref{VIA_Transport} 
\begin{equation}
	\mu_{ws}^0 = \sum_{\text{fam.}}\cbrak{3\mu_{q_L}+\mu_{l_L}}=\sum_{\text{fam.}}\cbrak{\frac{6}{T^2}}\cbrak{3~\frac{n_{q_L}}{\kappa_q}+\frac{n_{l_L}}{\kappa_l}}\simeq \frac{1}{2}\cbrak{\frac{6}{T^2}}\sum_\text{fam.}\cbrak{n_{q_L}+n_{l_L}}\equiv \frac{1}{2}\cbrak{\frac{6}{T^2}}n_L^0\,.
\end{equation}
Note that we applied the high-temperature expansion for the chemical
potentials to use the particle distribution functions $n_x$. The
factor $\nicefrac{T^2}{6}$ is absorbed in the transition rate. In the
second step, we used the zero-temperature statistical factors for
quarks and leptons, respectively. Since the strong sphaleron rate is
large compared to the electroweak sphaleron transition rate, the
involved chemical potentials of the left- and right-handed quark can
be related to each other,  
\begin{equation}
    0=\mu_{ss} \sim \sum_{\text{fam.}}\cbrak{n_{q_L}-n_{u_R}-n_{d_R}}\,.
\end{equation}
This allows us to relate the baryon asymmetry $n_B$ with the left-handed quark density as
\begin{equation}
	n_B = \frac{2}{3}\sum_\text{fam.}n_{q_L}\,.
\end{equation}
Since electroweak sphaleron transitions conserve $B-L$, the baryon asymmetry can be related with the left-handed lepton density
\begin{equation}
	n_B = \sum_\text{fam.} n_{l_L}\,,
\end{equation}
which finally allows us to formulate the differential equation for the
BAU 
\begin{equation}
    -v_W n_B^\prime - D n_B^{\prime\prime} = -\tilde{\Gamma}_{ws} \cbrak{\frac{3}{2}n_L^0 + \mathcal{R} n_B}\,,
    \label{eq::HO}
\end{equation}
with the SM relaxation term $\mathcal{R}=\nicefrac{15}{4}$. The \cref{eq::HO} can be solved numerically with the help of \ewbgBSMPT, where we use for the rescaled electroweak sphaleron transition rate 
\begin{equation}
	\tilde{\Gamma}_{ws} = 6\cdot \kappa \alpha_w^5 T_c\,,
\end{equation}
with the electroweak gauge coupling $\alpha_w$ and some numerical prefactor $\kappa$ of the order one. This value has a rather large theoretical error and is determined to be \cite{Moore:1997sn,Moore:1998mh}
\begin{equation}
	\kappa = 29\pm 6\,.
\end{equation}

\section{Numerical Analysis \label{sec:numerical}}
The main goal of our analysis is to investigate how the two different
approaches applied in the literature to compute the BAU compare to
each other and what are the crucial parameters that influence the
possible size of $\eta$. We furthermore want to understand how the
requirement of a strong first order EWPT combined with the strict
experimental constraints on the still allowed CP violation interacts 
with the goal to generate a BAU compatible with the observations. 
Before we present our results, however, we first give the details of our parameter scan.  

\subsection{Minimisation of the Effective Potential}
For the numerical determination of the strength of the phase
transition $\xi_c$, we use \texttt{BSMPT v2.2} \cite{Basler:2020nrq}
which extends \texttt{BSMPT} \cite{Basler:2018cwe} by the computation
of the electroweak baryogenesis in the C2HDM\footnote{Note also that
  in \texttt{BSMPT v2.2} an updated description of the numerical
  methods used in \texttt{BSMPT} is given. }, and we extend
the C2HDM parameter scan discussed in \cite{Basler:2019iuu}. The
search for parameter points that provide a strong first order EWPT, 
that are compatible with the Higgs data and that simultaneously produce the
correct amount of baryon asymmetry is a non-trivial task so that we
had to increase significantly the amount of scanned points compared to
the scan performed in \cite{Basler:2019iuu}. Since the numerical
determination of $\xi_c$ is not the main emphasis of this paper we
refer to \cite{Basler:2016obg,Basler:2019iuu,C2HDM} for
the detailed discussion of the impact of the requirement of a strong first order
EWPT on the collider phenomenology. 

\subsection{Constraints and Parameter Scan}
In the following we list the numerical values of the input parameters used in the
analysis where we focus on the C2HDM Type I (TI) and II (TII). In the parameter
scan, one of the neutral Higgs bosons, called $h$ in the following, is
required to have a mass of $m_h=125.09\gev$ \cite{HiggsAtlas} and
behave SM-like. The remaining two neutral Higgs bosons will be denoted as
$\hdown$ and $\hup$, where $\mdown<\mup$. We explicitly allow for all
three possible mass hierarchies 
\begin{subequations}
	\begin{align}
		\text{M}_{\text{I}}:\qquad   & m_h<\mdown<\mup\,, \\
		\text{M}_{\text{II}}:\qquad  & \mdown<m_h<\mup\,, \\
		\text{M}_{\text{III}}:\qquad & \mdown<\mup<m_h\,,
	\end{align}
\end{subequations}
The scan ranges for the input parameters of
the C2HDM TI are given in \cref{C2HDM::ParamT1} and for the C2HDM
TII in \cref{C2HDM::ParamT2}.
\begin{table}[b]
	\centering
	\begin{tabular}{c c c c c   }
		\toprule
		$m_{h}$  & $\mdown$               & $\mup$                 & $\mHc$                 & $\mbox{Re} m_{12}^2$              \\
		         &                        & in $\gev$              &                        & in $\gev^2$                 \\\midrule
		$125.09$ & $\left[30,1500\right]$ & $\left[30,1500\right]$ & $\left[30,1500\right]$ & $\left[10^{-3},10^5\right]$ \\\bottomrule
		$\alpha_1$  & $\alpha_2$ & $\alpha_3$ & $\tan\beta$\\\midrule
		$\left[-\frac{\pi}{2} , \frac{\pi}{2}\right)$ &  $\left[-\frac{\pi}{2} , \frac{\pi}{2}\right)$ &  $\left[-\frac{\pi}{2} , \frac{\pi}{2}\right)$ & $\left[0.8,20\right]$\\\bottomrule
	\end{tabular}
	\caption{Parameter ranges for the C2HDM TI input parameters used
		in {\tt ScannerS}.}
	\label{C2HDM::ParamT1}
	\vspace*{0.5cm}
\end{table}
\begin{table}[t]
	\centering
	\begin{tabular}{c c c c c   }
		\toprule
		$m_{h}$  & $\mdown$               & $\mup$                 & $\mHc$                  & $\mbox{Re} m_{12}^2$              \\
		         &                        & in $\gev$              &                         & in $\gev^2$                 \\\midrule
		$125.09$ & $\left[30,1500\right]$ & $\left[30,1500\right]$ & $\left[580,1500\right]$ & $\left[10^{-3},10^5\right]$ \\\bottomrule
		$\alpha_1$  & $\alpha_2$ & $\alpha_3$ & $\tan\beta$\\\midrule
		$\left[-\frac{\pi}{2} , \frac{\pi}{2}\right)$ &  $\left[-\frac{\pi}{2} , \frac{\pi}{2}\right)$ &  $\left[-\frac{\pi}{2} , \frac{\pi}{2}\right)$ & $\left[0.8,20\right]$\\\bottomrule
	\end{tabular}
	\caption{Parameter ranges for the C2HDM TII input parameters
		used in {\tt ScannerS}.}
	\label{C2HDM::ParamT2}
\end{table}
As for the remaining SM parameters, we use the fine structure constant taken at the $Z$ boson mass scale \cite{Agashe:2014kda,LHCHXSWG},
\begin{equation}
	\alpha_{\text{EM}}^{-1}(M_Z^2) = 128.962 \;,
\end{equation}
and the masses for the massive gauge bosons are chosen as \cite{Agashe:2014kda,LHCHXSWG}
\begin{equation}
	m_W = 80.385\gev \quad \text{and }\quad m_Z=91.1876\gev\,.
\end{equation}
The lepton masses are set to \cite{Agashe:2014kda,LHCHXSWG}
\begin{equation}
	m_e=0.511~\mathrm{MeV},\quad m_{\mu}=105.658~\mathrm{MeV},\quad m_{\tau} = 1.777\gev\,,
\end{equation}
and the light quark masses to \cite{LHCHXSWG}
\begin{equation}
	m_u = m_d = m_s = 100~\mathrm{MeV}\,.
\end{equation}
To be consistent with the CMS and ATLAS analyses, we take the on-shell top quark mass as \cite{LHCHXSWG,Dittmaier:2011ti}
\begin{equation}
	m_t = 172.5\gev\,
\end{equation}
and the recommended charm and bottom quark on-shell masses \cite{LHCHXSWG}
\begin{equation}
	m_c=1.51\gev \quad \text{and} \quad m_b = 4.92\gev\,.
\end{equation}
We choose the complex parametrization of the CKM matrix \cite{Chau:1984fp,Agashe:2014kda}
\begin{equation}
    V_{\text{CKM}} =
\begin{pmatrix}
    c_{12}c_{13} & s_{12}c_{13} & s_{13}\e^{-\ii \delta}\\
    -s_{12}c_{23}-c_{12}s_{23}s_{13}\e^{\ii\delta} & c_{12}c_{23}-s_{12}s_{23}s_{13} \e^{\ii \delta} & s_{23}c_{13}\\
    s_{12}s_{23}-c_{12}c_{23}s_{13}\e^{\ii\delta} & - c_{12}s_{23}-s_{12}c_{23}c_{13}\e^{\ii\delta} & c_{23}c_{13}
\end{pmatrix}\,,
  \end{equation}
where $s_{ij}=\sin\theta_{ij}$ and $c_{ij}=\cos\theta_{ij}$. The angles are given in terms of the Wolfenstein parameters
\begin{align}
	s_{12} = \lambda \,,\quad && s_{13}\e^{\ii \delta} = A \lambda^3\cbrak{\rho+\ii \eta} \,,\quad &&s_{23} = A \lambda^2 \,,
\end{align}
with \cite{Basler:2019iuu}
\begin{align}
	\lambda =0.22537\,,&& A =0.814 && \rho = 0.117 && \eta =  0.353\,.
\end{align}
Note that we take into account a complex phase $\delta$ in the CKM matrix as an
additional source for CP violation. The impact of the
complex CKM phase compared to that of the complex phase induced by the
VEV configuration is negligible, however. 
Finally, the electroweak VEV is set to
\begin{equation}
	v=1/\sqrt{\sqrt{2}G_F}=246.22\gev\,.
\end{equation}

The parameter points under investigation have to fulfil experimental
and theoretical constraints. For the generation of such parameter
points we use the {\tt C++} program {\tt ScannerS v2.0.0}
\cite{Coimbra:2013qq,Ferreira:2014dya,Muhlleitner:2016mzt,Muhlleitner:2020wwk}. \ScannerS
allows us to check for boundedness from below of the tree-level
potential and uses the tree-level discriminant of
\cite{Ivanov:2015nea} to ensure the electroweak vacuum to be the
global minimum at tree level. By using \texttt{BSMPT} it is also
possible to check for the NLO electroweak vacuum to be the global
minimum of the potential. Only parameter points providing a stable NLO
electroweak vacuum at zero temperature are taken into account for the
analysis. 
To be consistent with recent flavour constraints, we test for the compatibility
with $\mathcal{R}_b$ \cite{Haber:1999zh,Deschamps:2009rh} and
$B\rightarrow X_s \gamma $
\cite{Deschamps:2009rh,flavor1,flavor2,flavor3,charged580} in the
$m_{H^{\pm}}-\tan\beta$ plane. For the C2HDM TII, this implies that the
charged Higgs mass has to be above 580~GeV \cite{charged580} whereas
in the C2HDM TI this bound is much weaker and is strongly correlated
with $\tan\beta$. 
The compatibility with the Higgs measurements is taken into account by
{\tt ScannerS} through {\tt
  HiggsBounds}~\cite{Bechtle:2008jh,Bechtle:2011sb,Bechtle:2013wla}
and {\tt HiggsSignals}~\cite{Bechtle:2013xfa}. For the parameter scan
the versions {\tt HiggsBounds5.7.1} and {\tt HiggsSignals2.4.0} are
used. 
For the determination of the strength of 
the EWPT and the actual calculation of the BAU we use our new code
\texttt{BSMPT v2.2} \cite{Basler:2020nrq}. For this analysis, we chose in the
counterterm potential, {\it cf.}~Eq.~(\ref{eq:countertermpot}), $t_1=0$ and $t_2$ such
that for the counterterms $\delta 
\mbox{Im} \lambda_{6,7}$ we have $\delta \mbox{Im} \lambda_{6}=\delta
\mbox{Im} \lambda_{7}$.\footnote{We found that the results do not
  change if we set {\it e.g.}~$\delta \mbox{Im} \lambda_{6}=0$.} The
wall velocity which is an input parameter in {\tt BSMPT v2.2} is set as 
\begin{eqnarray}
v_W = 0.1 \;.
\end{eqnarray}

Altogether we found 186 parameter points that fulfill all experimental
and theoretical constraints and simultaneously provide a strong first
order EWPT. We start the analysis with the discussion of the
additional counterterms, followed by the discussion of the found BAUs
in the {\tt FH} and {\tt VIA} approaches.  
Afterwards, we investigate the impact of the chosen bubble wall
velocity and finally we take the full sample of parameter 
points in our analysis and compare both approaches applied in the
computation of the BAU.
\subsection{Discussion of the Counterterms}
\label{sec::counterterms}
\begin{figure}[h]
  \centering
  \includegraphics[width=0.6\textwidth]{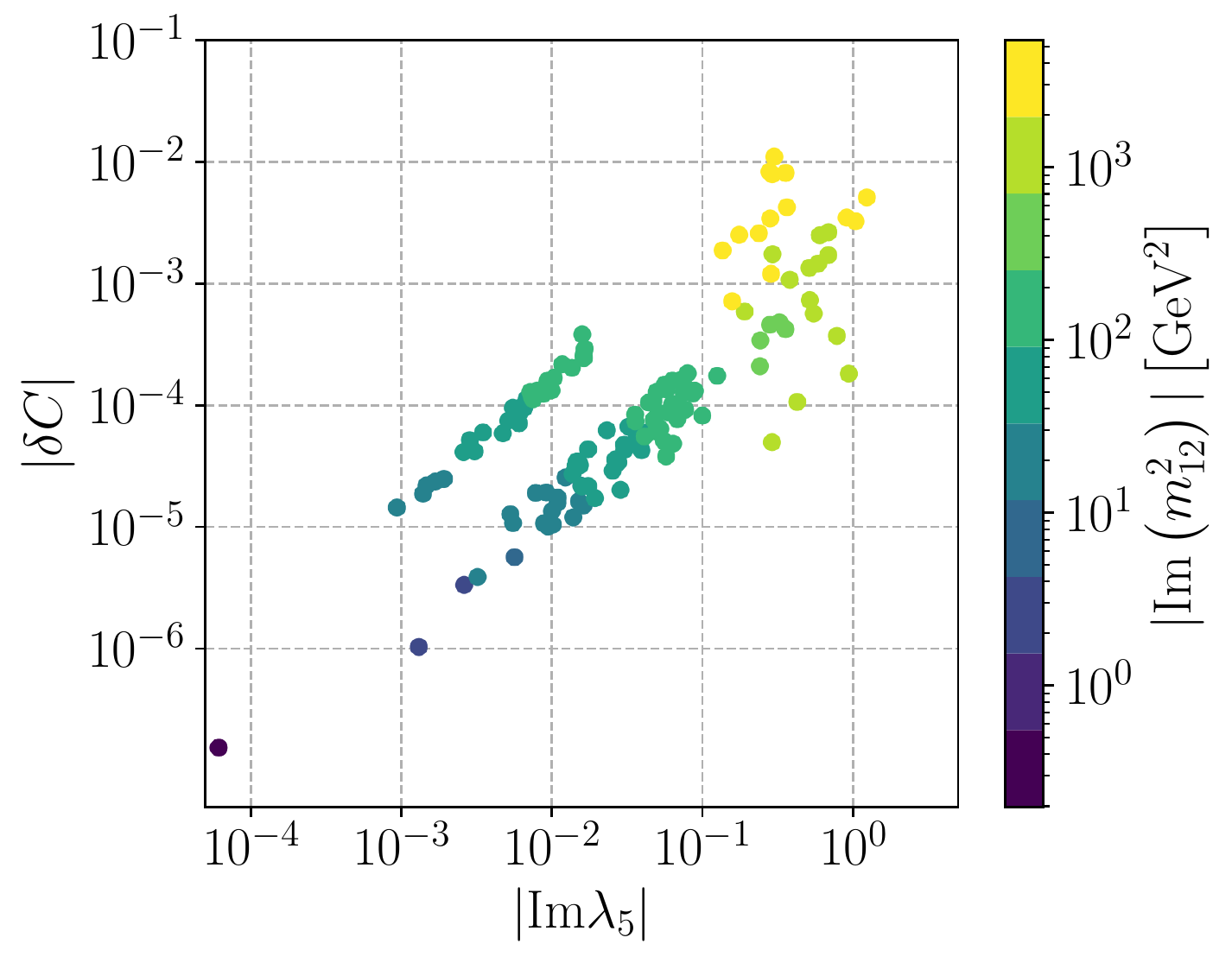}
\caption{The absolute value of the counterterm $\delta C \equiv \delta \mbox{Im}
  \lambda_6 = \delta \mbox{Im} \lambda_7$ for all points passing our
  constraints with $\xi_c\ge 1$ as function of $|\mbox{Im}
  \lambda_5|$. The color code denotes the size of
  $|\mbox{Im}(m_{12}^2)|$ in GeV$^2$. \label{fig:ctsize}}
\end{figure}

We start by discussing the influence of the radiatively generated
counterterms $\delta \mbox{Im} \lambda_{6,7}$. In
Fig.~\ref{fig:ctsize} we plot the size of $\delta C \equiv \delta \mbox{Im}
  \lambda_6 = \delta \mbox{Im} \lambda_7$ for all points of our
  parameter scan that fulfill the described constraints and have
  $\xi_c \ge 1$, as a function of $|\mbox{Im} \lambda_5|$. The color
  code indicates the size of $|\mbox{Im}(m_{12}^2)|$ in GeV$^2$. The
  plot clearly shows that the new counterterm scales with the
  imaginary parts of $\lambda_5$ and $m_{12}^2$ and thus with the
  complex phase allowing for explicit CP-violation. In the
  CP-conserving limit these imaginary parts would vanish so that no flavour
  violation is generated radiatively and hence no such counterterms
  would be required.\footnote{For a recent discussion of the interplay
    of CP violation and $\mathbb{Z}_2$ breaking under a 2-loop renormalization
    group analysis, see \cite{Oredsson:2019mni}. While CP violation easily spreads
    across the Higgs and Yukawa sectors during renormalizaton group
    evolution when $\mathbb{Z}_2$ is broken, induced flavour-changing
    neutral currents (FCNCs) are not very large for points compatible
    with the EDMs.} We 
  also checked different choices of the free parameter $t_2$ relevant
  for the determination of $\delta \mbox{Im} \lambda_{6,7}$, {\it
    cf.}~Eqs.~(\ref{eq:glgb})-(\ref{eq:glge}), by setting {\it
    e.g.}~$\delta \mbox{Im} \lambda_6=0$, and found that the results
  did not change. The default setting of {\tt BSMPT v2.2} is hence
  $t_1=0$ and $t_2$ such that $\delta \mbox{Im}
  \lambda_6 = \delta \mbox{Im} \lambda_7$, which can be changed,
  however, by the user if desired.
We finally remark that since our renormalization scheme is set up such 
that the Higgs masses 
and mixing remain at their tree-level values no dangerous 
FCNCs are induced at tree level through the
loop-corrected potential.\footnote{Since possible FCNCs are induced
  only at loop-level and the new counterterm contributions are found
  to be small we expect the impact of the loop-induced FCNCs to be
  sufficiently small to be compatible with experiment. Since our focus
  here is on the investigation if in our model it is at all possible
  to generate a BAU large enough to be 
  compatible with experiment we leave the detailed analysis of this
  aspect for future work.}

\subsection{The Amount of Generated Baryon Asymmetry}
\label{sec::overallsize}
\begin{figure}[h]
  \centering
  \includegraphics[width=0.6\textwidth]{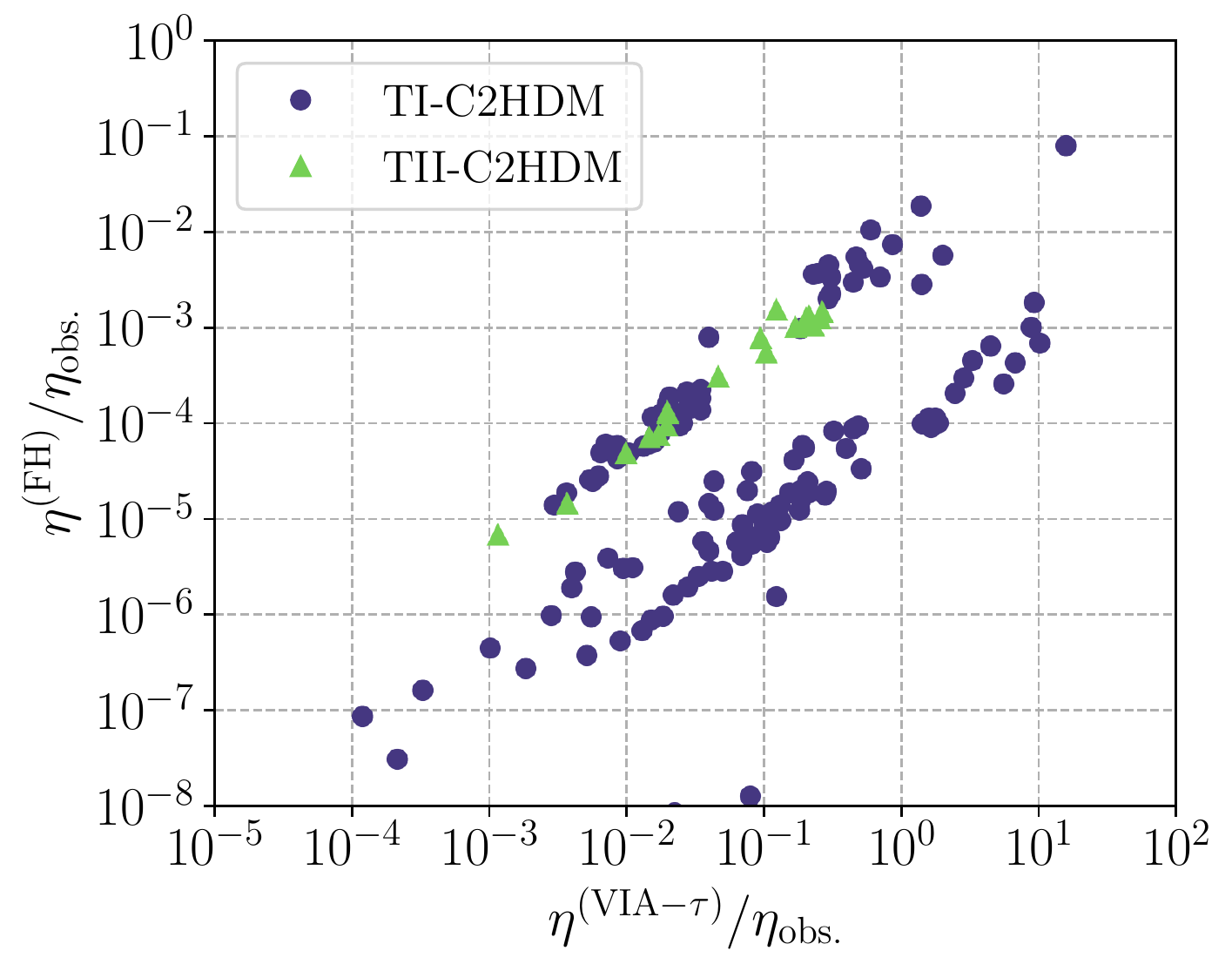}
\caption{BAUs in the {\tt FH} approach versus the {\tt VIA}
approach, including $t$, $b$ and $\tau$ contributions in the latter,
both normalized to the observed value. Results for the C2HDM TI are
shown as violet points, those for the for the C2HDM as green
triangles.
\label{fig:barasym}}
\end{figure}

In Fig.~\ref{fig:barasym} we show the computed generated baryon
asymmetry $\eta$ for our allowed scan points in type 1, denoted TI,
(violet points) and type 2, denoted TII, (green triangles) in the {\tt FH} and in the {\tt VIA}
approach, both normalized to the observed baryon asymmetry
$\eta_{\text{obs}}$. In the {\tt VIA} approach the massive $t$, $b$
and $\tau$ contributions are taken into account. The impact of the
different inclusions will be discussed later. We first remark that
both approaches for the derivation of the quantum transport equations
are correlated in the sense that they predict the largest BAU for the
same parameter points. However, the {\tt VIA} method predicts BAU
values that are two to three orders of magnitude larger than those obtained
in the {\tt FH} method. This issue has been discussed in the
literature \cite{Cline:2020jre} leading to some criticism with respect to the
validity of the approximations made in the {\tt VIA} method. It was
argued that the expansion applied in the derivation of the source term
for the top quark might break down because of the large top quark mass
\cite{Cline:2020jre,Postma:2019scv}. This might be the reason why it
is possible to generate such large values for the BAU. Hence, the
{\tt VIA} method is able to predict a BAU that is compatible with the
observed value for the parameter points passing the constraints of our
scan while it is not possible in the {\tt FH}
approximation. Actually, we did not find any benchmark point that
provides the sufficient amount of BAU together with a small bubble wall
velocity within the {\tt FH} approach.  \s

When we compare both types of C2HDM we see that in type 2 we have a
constant ratio between both approaches while type 1 shows a stronger
difference in the {\tt FH} and {\tt VIA} results. Overall, however,
the results in both C2HDM types are rather similar so that in the following
we will discuss both types of C2HDM together. 
\s

\subsection{Dependence on the Bubble Wall Velocity}
\label{sec::velocity}
\begin{figure}[h]
  \centering
  \includegraphics[width=0.49\textwidth]{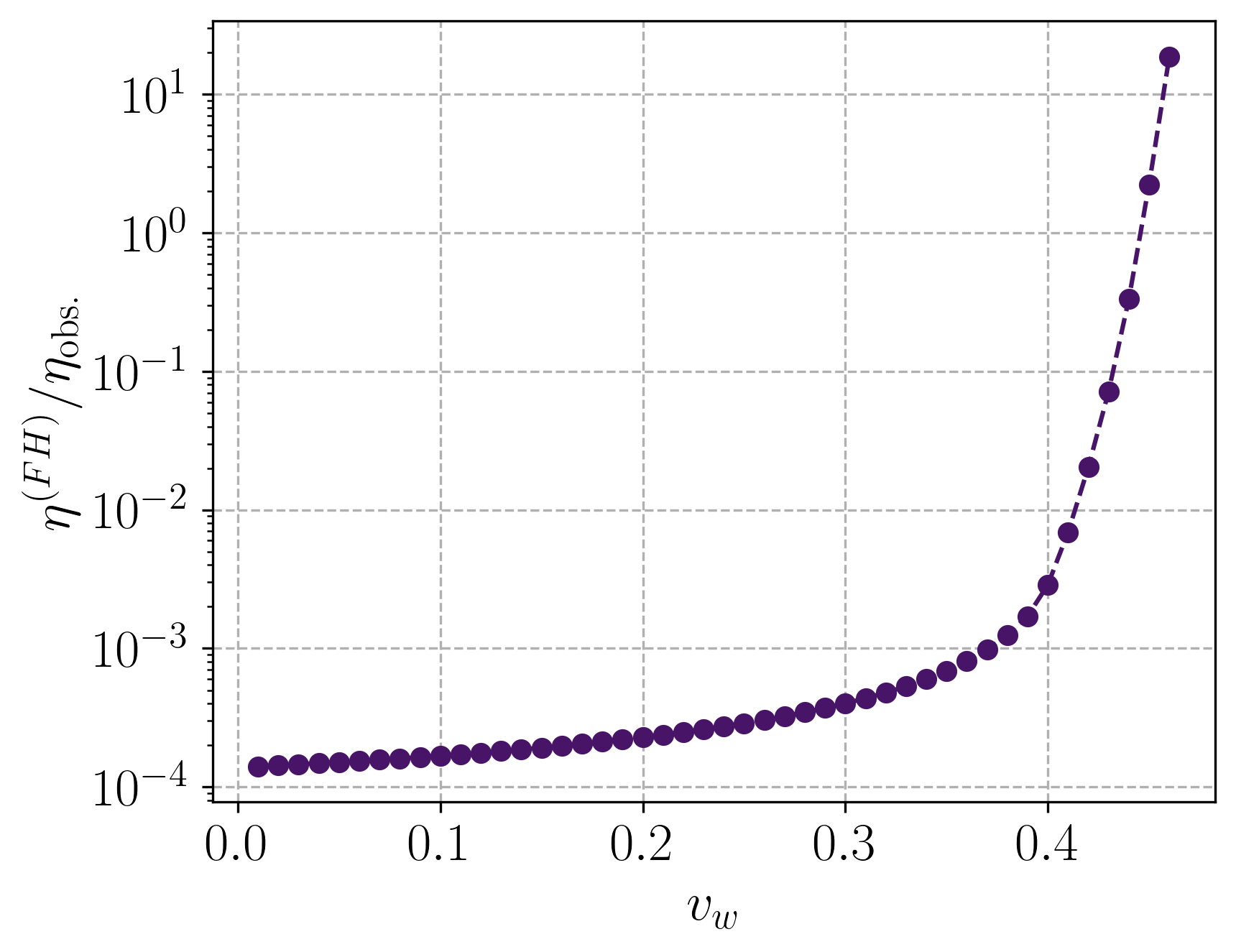}
  \includegraphics[width=0.49\textwidth]{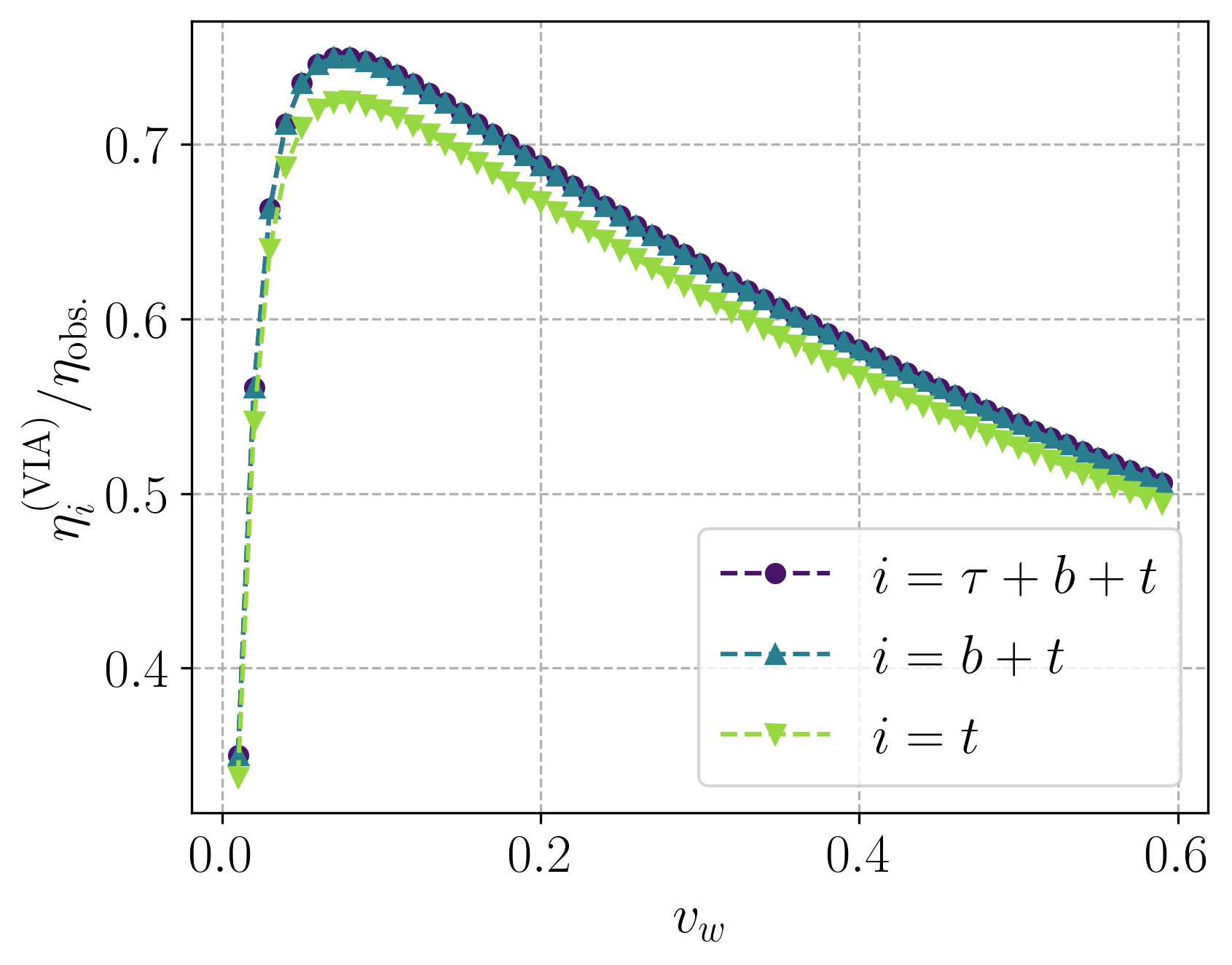}
\caption{{\tt BMPI}: BAU normalized to the observed value obtained in the {\tt FH}
approach (left) and in the {\tt VIA} approach (right) as function of
the wall velocity $v_W$. Results are shown in the {\tt VIA} approach
for the inclusion of the $t$, $b$ and $\tau$ contributions (violet
points), of the $t$ and $b$ contributions (blue triangles), and of the $t$
contributions only (green triangles).} 
\label{fig:velodep}
\end{figure}

As stated above, we set $v_W=0.1$ in our scans. Here now, we want to
discuss the dependence of the BAU on $v_W$ in both approaches. For
this we choose a specific benchmark point out of our sample of allowed
scan points. The input parameters for this point, called \texttt{BMPI} in the
following, are listed in Tab.~\ref{tab::BMP}. For this point we have a critical VEV and temperature of 235.26 GeV and 166.06 GeV,
respectively, and hence $\xi_c=1.42$. The complex phase of the top
quark mass is $\theta_t = 0.074$, the wall thickness $L_W=0.154$ and
$L_W T_c = 25.61$~GeV$\gg 1$. Starting from this
benchmark point we vary $v_W$ while keeping all parameters fixed and
compute the corresponding BAU.
\begin{table}[b]
	\centering
\begin{tabular}{c c c c c c c c}
		\toprule
		                $m_{h}~\left[\text{GeV}\right]$ & $m_{\hdown}~\left[\text{GeV}\right]$ & $m_{\hup}~\left[\text{GeV}\right]$ & $\mHc~\left[\text{GeV}\right]$ & $\tan\beta$ & $\alpha_1$ & $\alpha_2$ & $\alpha_3$ \\\midrule
		125.09  & 76.78 & 128.95  &  165.08 & 11.96 & -0.072 & 0.140 & 0.248
\\\midrule
	$\lambda_1$ & $\lambda_2$ & $\lambda_3$ & $\lambda_4$  & $\Re\lambda_5$ & $\Im\lambda_5$ & $\Re m_{12}^2~\left[\text{GeV}^2\right]$
& $\mbox{Im} m_{12}^2~\left[\text{GeV}^2\right]$
\\\midrule
		2.651 & 0.259 & 0.959 & -0.545 & -0.186 &  0.049 & 421.571 & 124.083
		\\\bottomrule
	\end{tabular}
	\caption{Input parameters of the benchmark point \texttt{BMPI} discussed in
          \cref{sec::velocity}: The parameter point is defined for the C2HDM TI. }
	\label{tab::BMP}
\end{table}
The result is shown in \cref{fig:velodep} which depicts the BAU in
the {\tt FH} approach (left) and in the {\tt VIA} approach right, both
normalized to the observed BAU as a function of $v_W$. In the {\tt
  VIA} approach we show results for the case where the massive $t$, $b$ and
$\tau$ contributions are taken into account in the transport equations
(violet points), where only $t$ and $b$ are included (blue triangles),
and with $t$ contributions solely (green triangles). \s

As can be inferred from the left plot, the {\tt FH} approach exhibits
a mild dependence on $v_W$ for small $v_W$. The {\tt FH} approach uses an explicit
expansion for small wall velocities and is only valid in this
regime. The mild dependence on $v_W$ ensures that the choice of the
input value for $v_W$ does not impact the resulting BAU
significantly. If the bubble wall velocity approaches the plasma sound
speed $v_W \approx 1/\sqrt{3}$, however, the BAU predicted in the {\tt
FH} approach, diverges. Recently it was found by the authors
of~\cite{Cline:2020jre} by re-deriving the fluid equations without making the
approximation of small $v_W$ that the sound speed barrier can safely
be crossed. In this context, also some mistakes in the previous
derivation of the {\tt FH} approach were pointed out. Their numerical
comparison of both old and new results showed that they agree for
small wall velocities and deviate by less than 30\% for $v_W=0.1$ in
the predicted BAU. The new approach will be implemented in the next
upgrades of {\tt BSMPT v2.2}. Apart from the steep fall for very small
velocities, the {\tt VIA} method shows a similarly mild dependence
on $v_W$ as the {\tt FH} approach. The {\tt VIA} method does not apply
an expansion in small $v_W$, but assumes small velocities, so that the
choice of $v_W=0.1$ is reasonable.  

\subsection{Wall Thickness and Mass Scale}
\label{sec::wall}
\begin{figure}[h]
  \centering
  \includegraphics[width=0.6\textwidth]{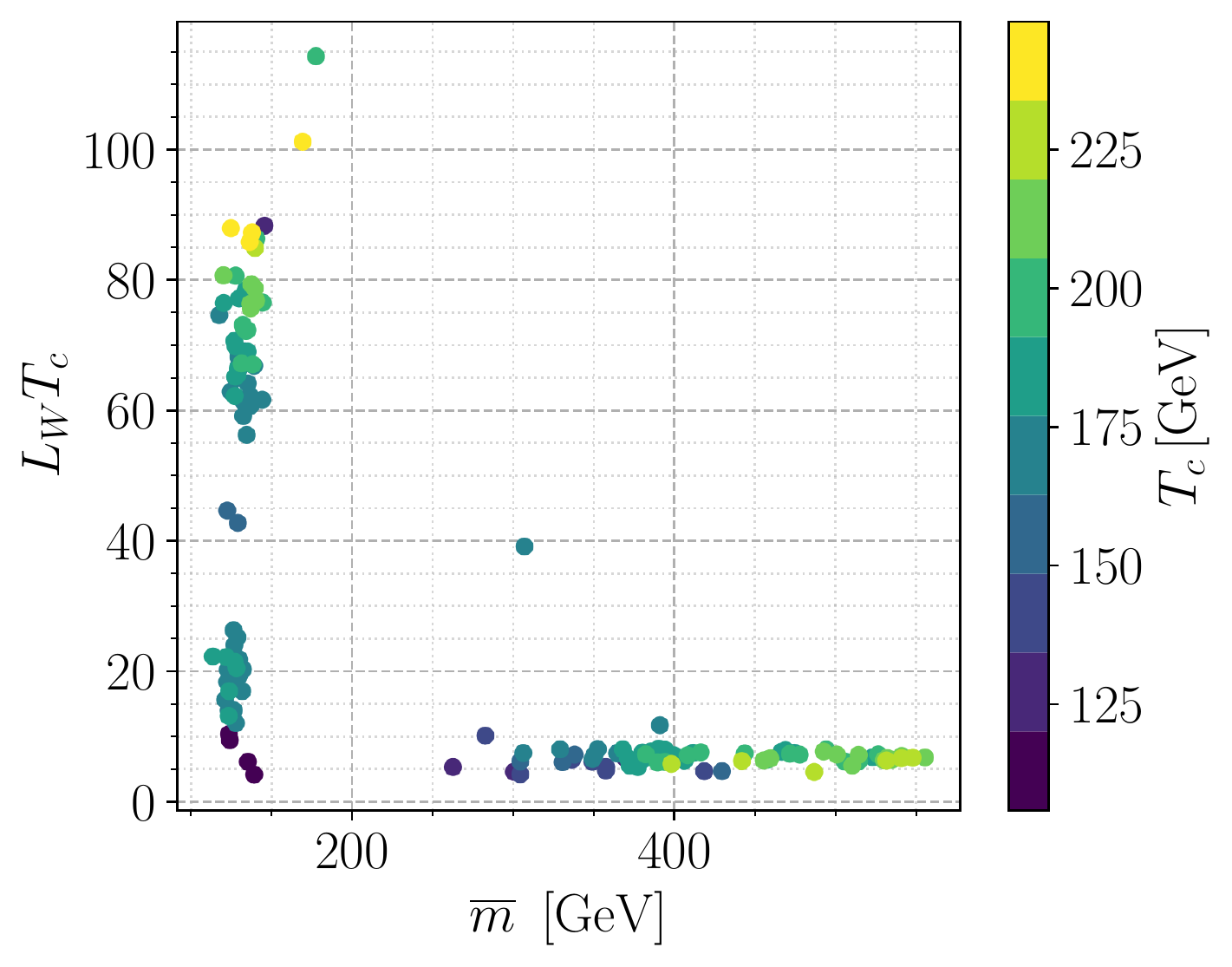}
\caption{$L_W T_c$ versus the average mass scale $\overline{m}$
  (definition, see text) for the allowed C2HDM TI and TII parameter
  points. The color
  code denotes the critical temperature $T_c$.
}
  \label{fig:LWM}
\end{figure}

As discussed in Sec.~\ref{sec:baryocalc}, the {\tt FH} ansatz works
for thick bubble walls. With the typical particle wavelength in the
plasma given by the inverse temperature $T^{-1}$ this implies the requirement
\beq
1 \ll L_W T_c \;.
\eeq
In Fig.~\ref{fig:LWM}, we see for the allowed C2HDM TI and
TII points the values of $L_W T_c$ as a function of the average mass
scale
\beq
\overline{m} = \frac{1}{4} \left( \sum_{i_=1}^3 m_{H_i} + m_{H^\pm} \right)
\;.
\eeq 
The color code denotes the values of the critical temperature
$T_c$. Apart from one outlier, there are two distinct regions in the
plot, given by small average mass values with $4 \lsim L_W T_c \lsim
115$ on the one hand and small $L_W T_c$ with $260 \mbox{ GeV } \lsim \overline{m}
\lsim 560 \mbox{ GeV}$ on the other hand. In both branches higher
$T_c$ are realized towards the upper end of each
branch. As can be clearly inferred
from the plot, large values of $L_W T_c$ are only realized for an
overall light mass spectrum.

\subsection{Scaling Behaviour of Both Approaches}
\label{sec::scaling}
Important parameters for successful baryogenesis and for the
approaches used in the computation are the complex phase $\theta_t$ of the top
quark mass, the strength of the phase transition $\xi_c$, and $L_W T_c$. 
The phase $\theta_t$ indicates the amount of CP violation which is required for
electroweak baryogenesis. The source terms in the transport equations
are proportional to the phase factor. The strength $\xi_c$ of the phase
transition can be viewed as a parameter describing the dynamics of the
phase transition. Its importance has been discussed in the literature
(see {\it e.g.}~\cite{Fromme:2006cm}). A stronger EWPT, {\it
  i.e.}~larger $\xi_c$, is expected to produce more BAU. The bubble
wall thickness times the critical temperature, $L_W T_c$, is required
to be large for the {\tt FH} method to be applicable. The wall
thickness $L_W$ itself is used for the parametrization of the bubble
wall profile and can be understood as a parameter describing the state
of the bubble. The bubble wall dynamics is given by the wall velocity
$v_W$, which we have set, however, to a fixed value, $v_W=0.1$, for
all of our parameter points. \s 
\begin{figure}[h]
  \centering
  \subfigure[]{\includegraphics[width=0.49\textwidth]{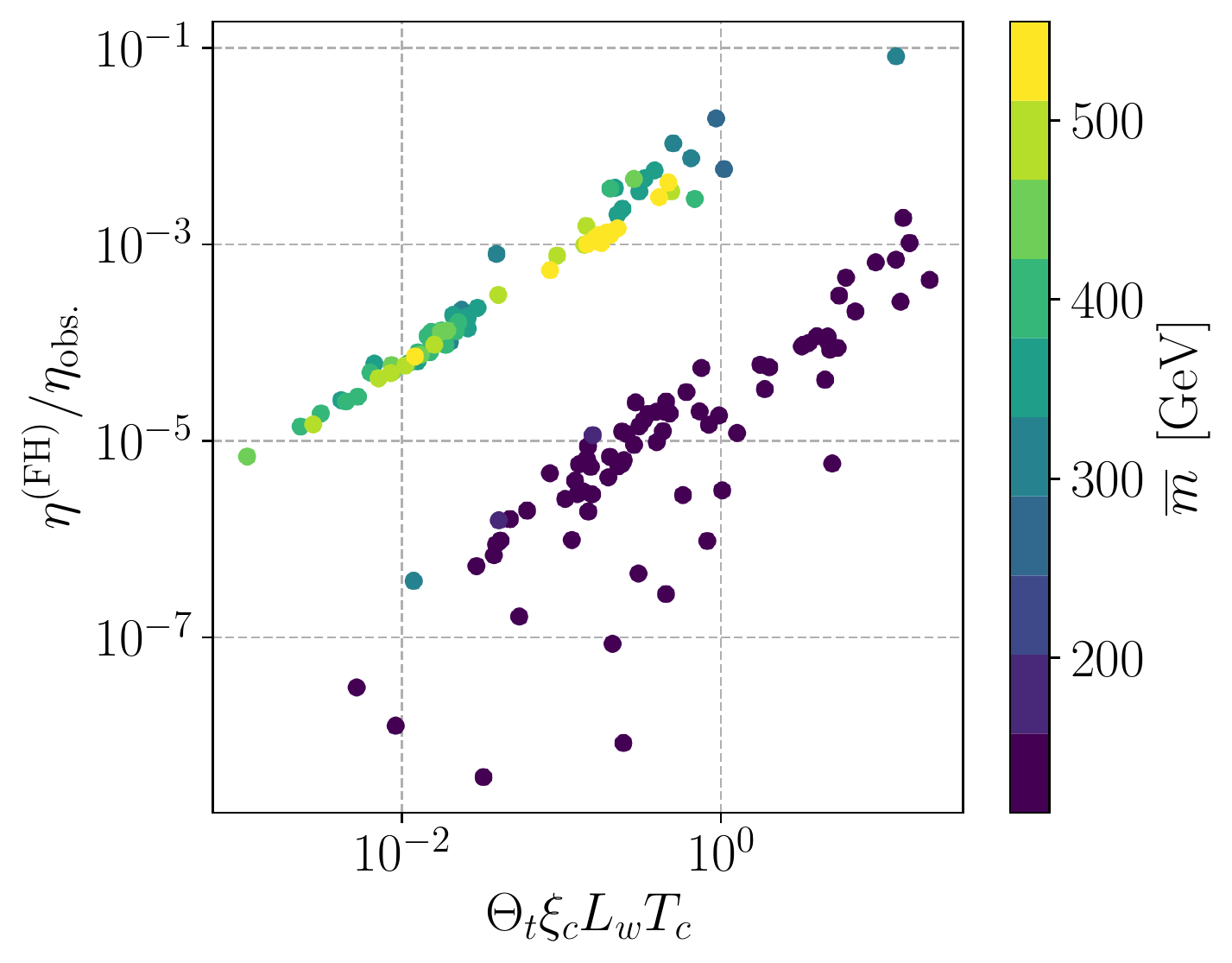}}
  \subfigure[]{\includegraphics[width=0.49\textwidth]{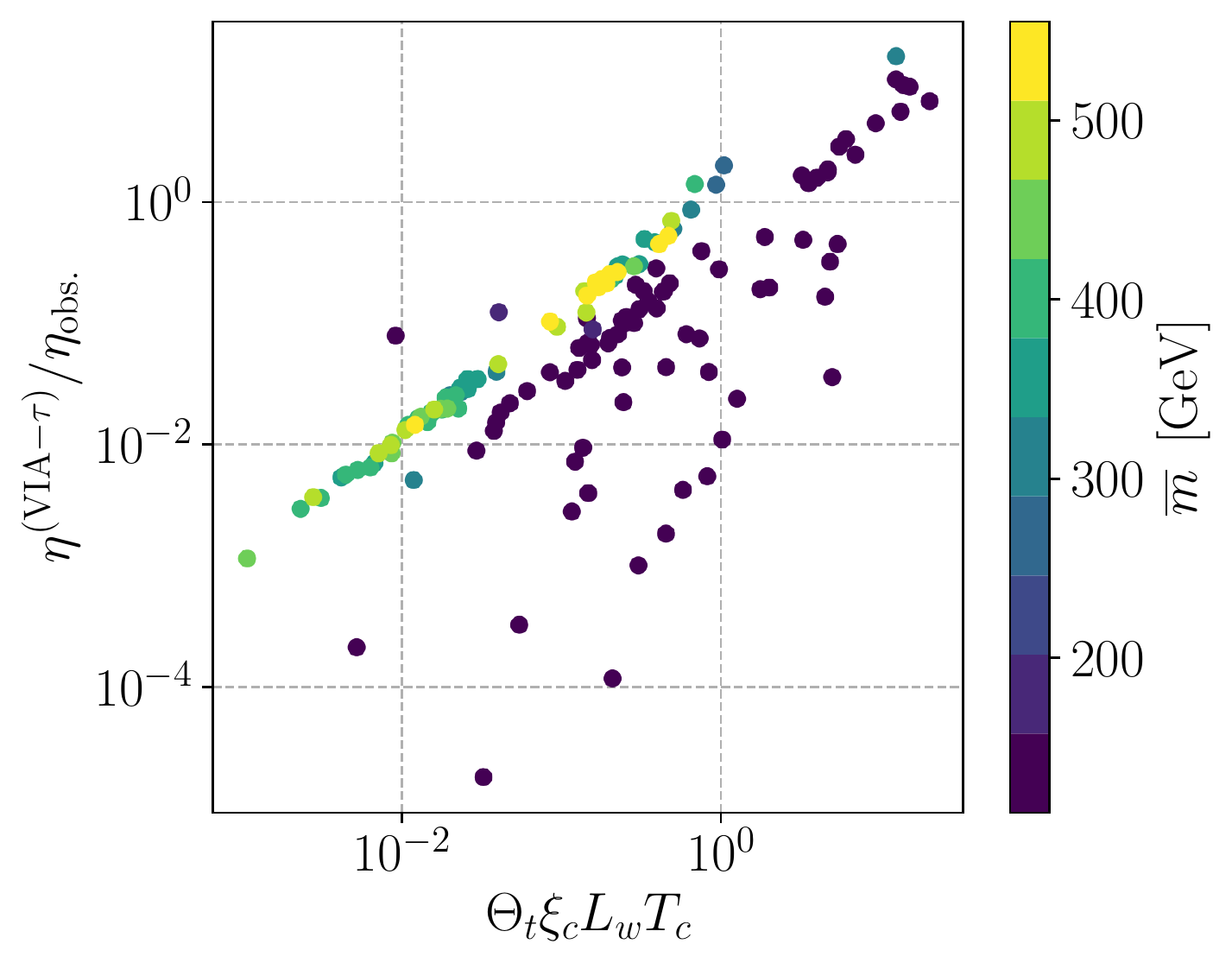}}
\caption{Normalized BAU in the {\tt FH} approach (left) and the {\tt
    VIA}$-\tau$ approach (right) as a function of the tuning parameter
combination $(\theta_t \xi_c L_W T_c)$ for the allowed parameter
points. The color code indicates the average mass scale $\overline{m}$. \label{fig:scaling}}
\end{figure}

In Fig.~\ref{fig:scaling} the normalized BAU is shown for our allowed
C2HDM TI and TII points as function of the tuning parameter
combination $(\theta_t \xi_c L_W T_c)$ for the {\tt FH} approach (left)
and the {\tt VIA}$-\tau$ approach, {\it i.e.}~including massive $t$,
$b$ and $\tau$ contributions, (right). The color code indicates the
size of the average mass scale $\overline{m}$. In both approaches the BAU
clearly increases with rising $\theta_t \xi_c L_W T_c$. 
We also see that the maximum values of BAU are obtained for a larger
average mass scale $\overline{m}$. Thus the {\tt VIA}$-\tau$ approach
reaches BAU values around the measured one for an average mass scale
above around 350~GeV. Also the {\tt FH} approach moves closer to the
measured value for a mass spectrum above 350~GeV, remains, however,
below the observed value. \s

The insights that we have gained so far allow us to discuss in more
detail what are the limiting factors in obtaining a large enough
BAU. A pre-requisite for successful BAU is a $\xi_c$ above one. The
influence of $\xi_c$ on $\eta$ should not be too strong here, as all
$\xi_c$ values that we could obtain in accordance with the applied
constraints range only between 1 and at most 1.95.
We need a large CP-violating phase for sufficient generation of a
baryon-antibaryon asymmetry. The CP-violating phase, however, is
severely constrained by the EDM measurements. Furthermore, an overall
heavier spectrum is advantageous for the amount of BAU as we have just
seen. On the other hand a strong first order EWPT favors a Higgs mass
spectrum where the Higgs bosons are close to each other in the
intermediate mass range \cite{Basler:2019iuu,C2HDM} and hence mix
strongly. Constraints from the oblique 
$S,T,U$ parameters force the charged Higgs mass to be degenerate with
one of the neutral Higgs bosons and constrain large mass differences
between Higgs states that are considerably mixed so that scenarios with large
mass gaps cannot be 
realized. Moreover, we have seen that a large value of $L_W T_c$ as
required in the {\tt FH} ansatz is only realized for a light mass
spectrum. These contradictory requirements on the overall Higgs mass
spectrum explain why it is difficult to reach large enough BAU. \s

\begin{figure}[t]
  \centering
  \subfigure[]{\includegraphics[width=0.49\textwidth]{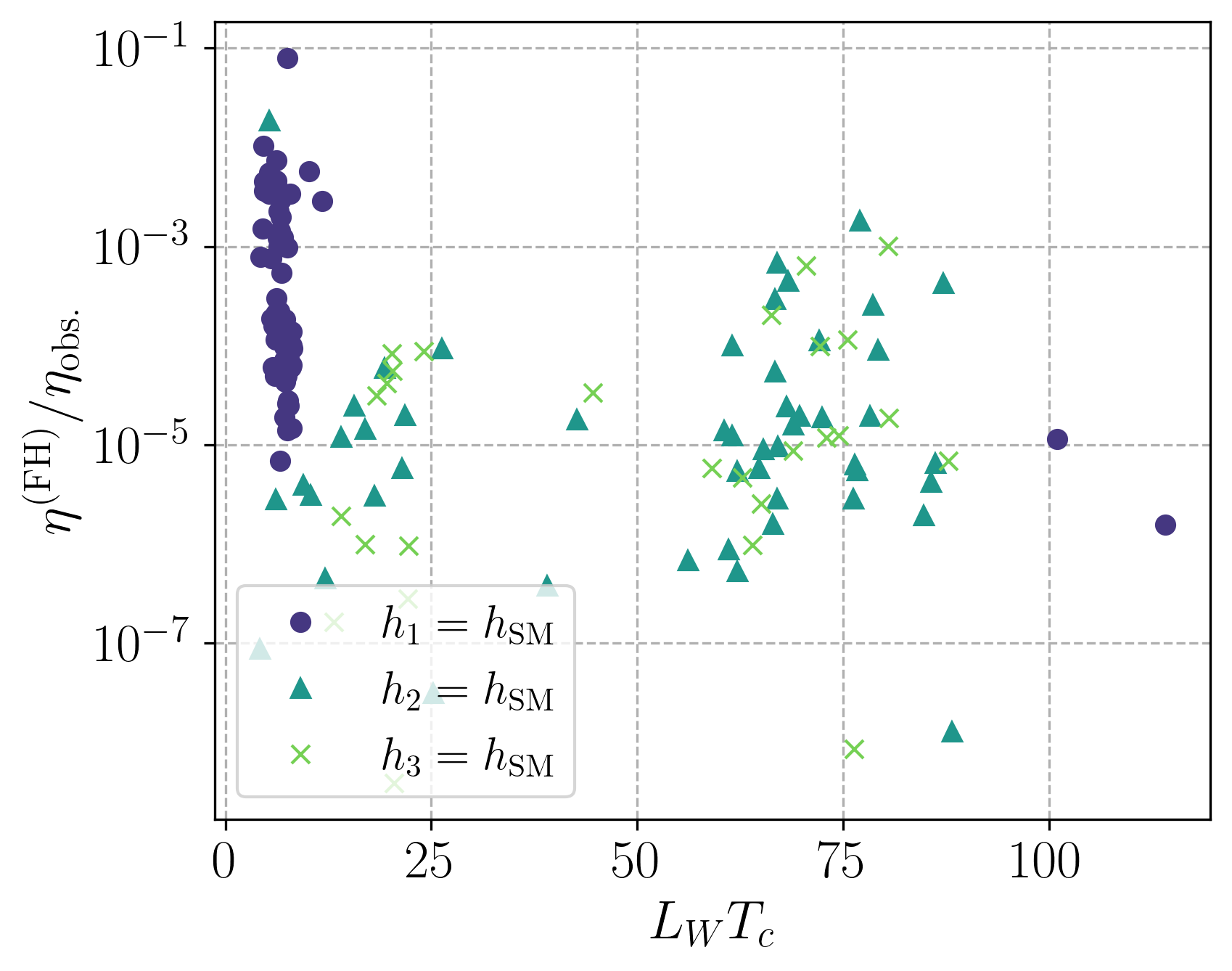}}
  \subfigure[]{\includegraphics[width=0.49\textwidth]{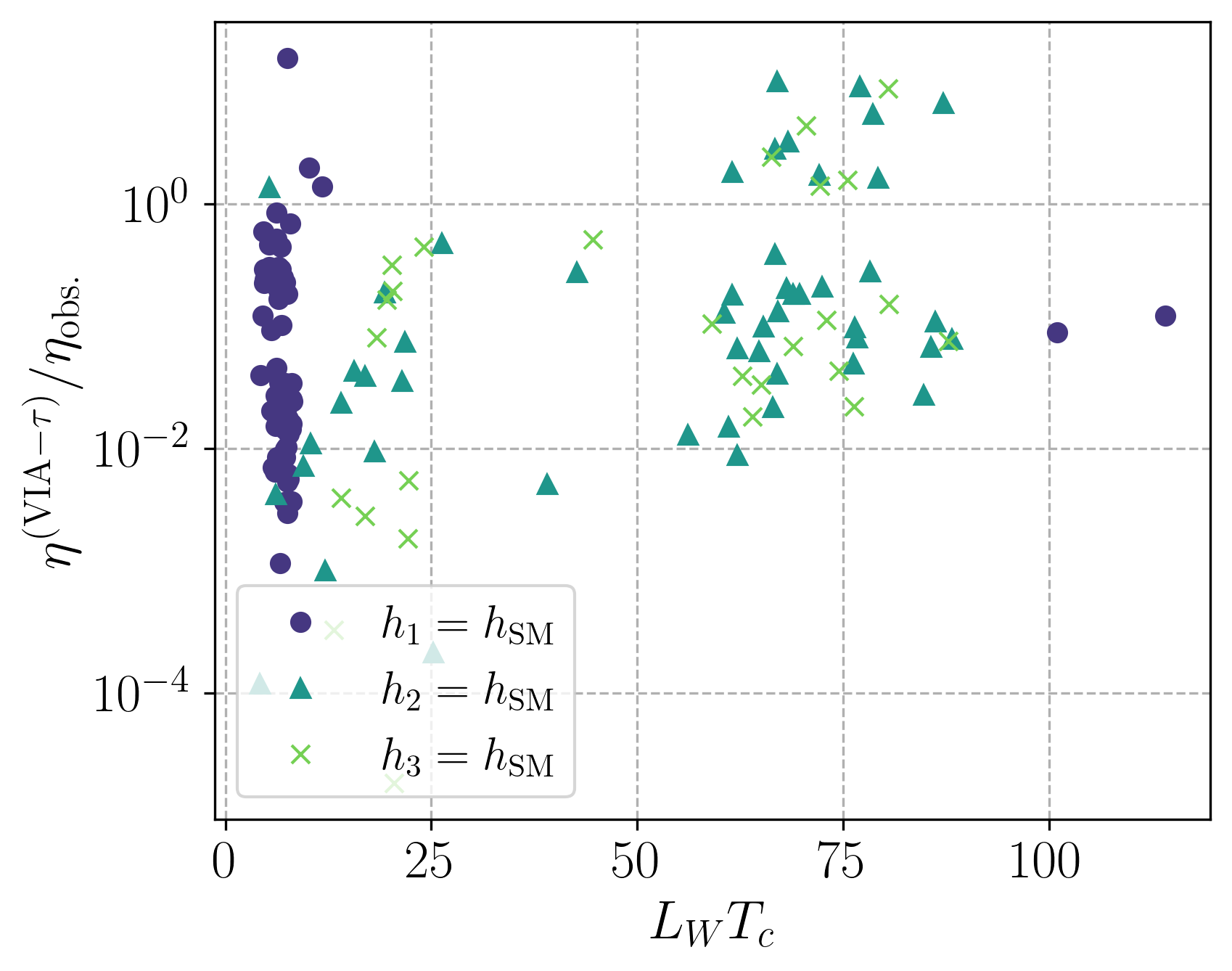}}
\caption{Normalized BAU in the {\tt FH} approach (left) and the {\tt
    VIA}$-\tau$ approach (right) as a function of $L_W T_c$ for the allowed parameter
points and $h_1$ being the SM-like Higgs boson $h_{\text{SM}}$ (violet
dots), $h_2$ being SM-like (dark-green triangles), and
$h_3=h_{\text{SM}}$ (green crosses). \label{fig:lwtc}}
\end{figure}
\begin{figure}[t]
  \centering
  \includegraphics[width=0.5\textwidth]{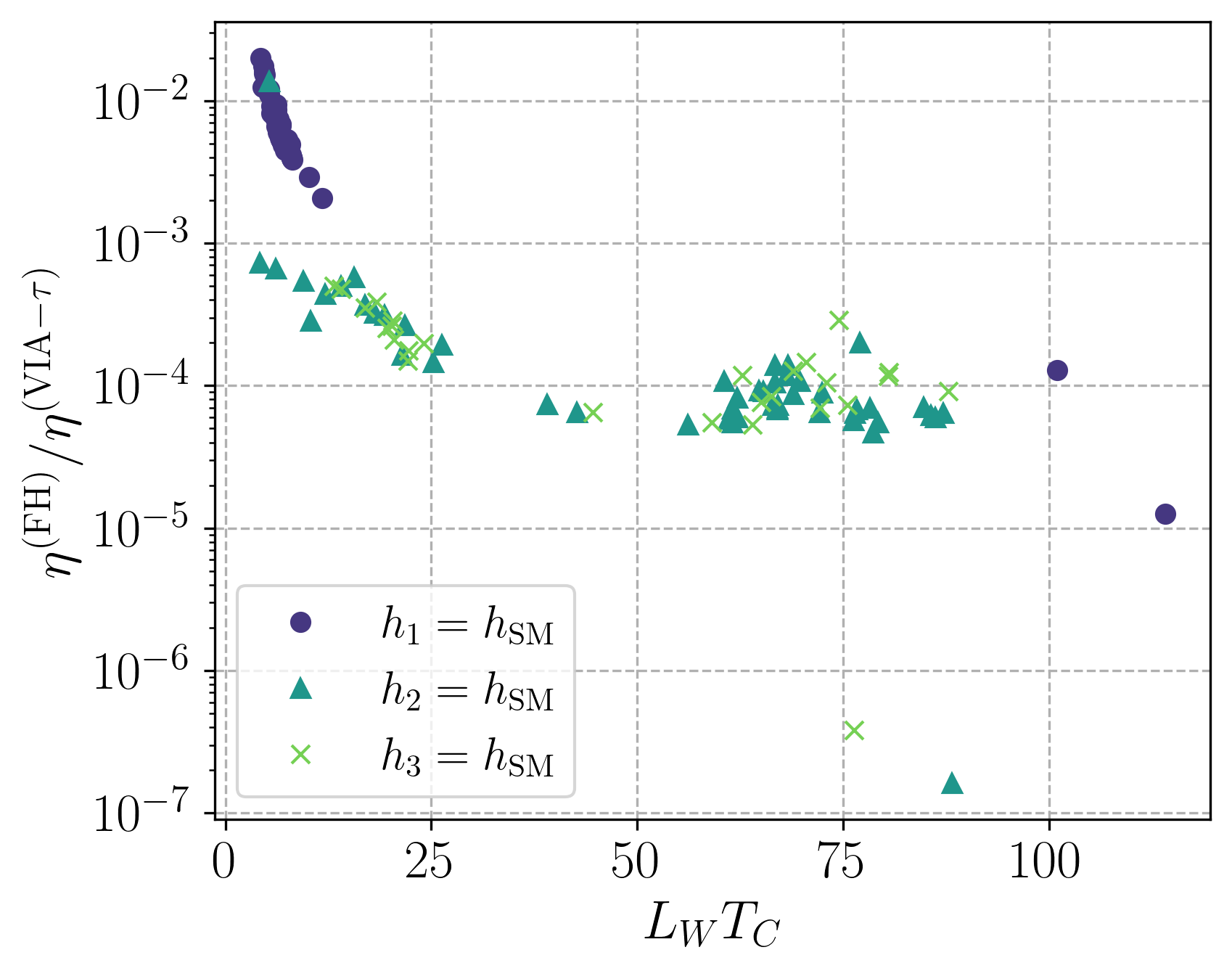}
\caption{Ratio of the BAU in the {\tt FH} approach and in the {\tt
    VIA}$-\tau$ approach as a function of $L_W T_c$ for the allowed parameter
points and $h_1$ being the SM-like Higgs boson $h_{\text{SM}}$ (violet
dots), $h_2$ being SM-like (dark-green triangles), and
$h_3=h_{\text{SM}}$ (green crosses). \label{fig:ratiofhvia}}
\end{figure}
\begin{figure}[h]
  \centering
  \subfigure[]{\includegraphics[width=0.49\textwidth]{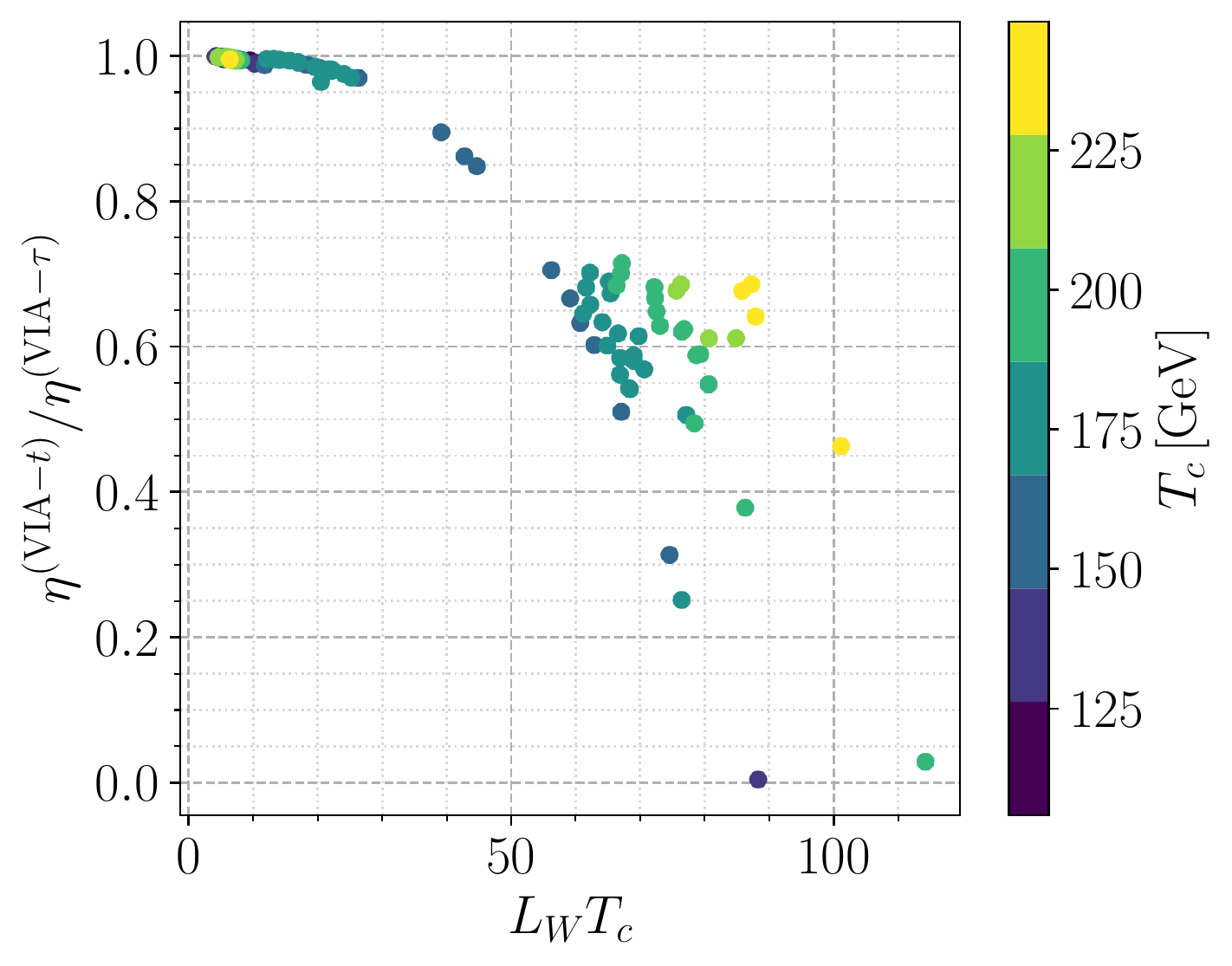}}
  \subfigure[]{\includegraphics[width=0.49\textwidth]{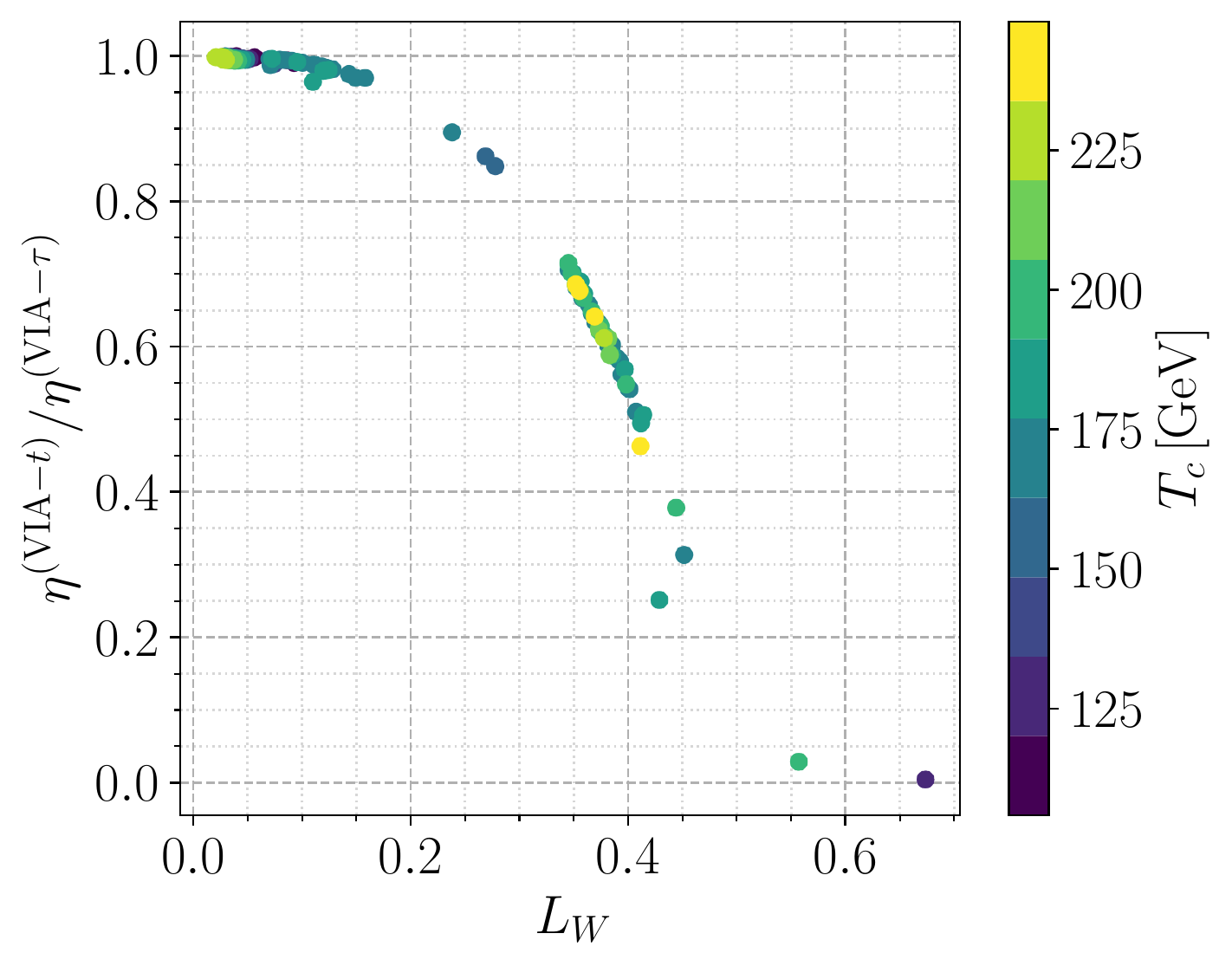}}
\caption{Ratio between the BAU computed in the {\tt VIA}$-t$ approach
  and the {\tt VIA}$-\tau$ approach as a function of $L_W T_c$ (left)
  and $L_W$ (right) for all allowed parameter points. The color code
  indicates the critical temperature $T_c$. \label{fig:addferm}}
\end{figure} 
The only parameter to be discussed with respect to the amount of BAU
is $L_W T_c$ which we will do next. Figure \ref{fig:lwtc} shows the
dependence of the BAU in the {\tt FH} approach (left) and the {\tt
  VIA}$-\tau$ approach (right) normalized to the observed value as a
function of $L_W T_c$ for the allowed parameter points. The color and
shape code indicates which of the three neutral Higgs bosons is the
SM-like one. For the {\tt VIA}$-\tau$ approach there is no clear
correlation between the amount of BAU and the size of $L_W T_c$. The
{\tt FH} method shows an increase of $\eta$ for small $L_W
T_c$ and $h_1$ being the SM-like Higgs boson, {\it i.e.} in a region
where we approach the lower limit $L_W T_c > 1$ required for the
application of the {\tt FH} method. For larger $L_W T_c$ the dependence on
$L_W T_c$ is similar to the {\tt VIA}$-\tau$ approach indicating an
agreement in the diffusion description of both approaches with the
{\tt FH} approach predicting less BAU, however. 
This can also be inferred from Fig.~\ref{fig:ratiofhvia} which
  shows the ratio of the BAU in the {\tt FH} approach and in the {\tt
    VIA}$-\tau$ approach as a function of $L_W T_c$ for the allowed
  parameter points. The ratio steeply increases for small values of 
  $L_W T_c$ to about $2 \times 10^{-2}$ and becomes rather constant,
  with ratios around $10^{-4}$, for larger values of $L_W T_c$.
While many points with
normal mass ordering appear for small $L_W T_c$ we still find
parameter points, where $L_W T_c$ is large for this mass ordering and
where both methods can be applied. 

\subsection{The Effect of Additional Fermions}
\label{sec::addfermions}
 In Fig.~\ref{fig:addferm} we display, for the allowed points, the
ratio of the baryon asymmetry computed in the {\tt VIA}$-t$ approach
where only the top quark has been included and the one computed in the
{\tt VIA}$-\tau$ approach with the top, bottom and $\tau$ contributions
included in the transport equations. The left plot shows the ratio as
function of $L_W T_c$ and the right one shows the ratio as function of
$L_W$. The color code indicates the critical temperature. Our investigations
show that the inclusion of $\tau$ has little effect on $\eta$. The
additional inclusion of the bottom contribution, however, has a
significant increasing effect on $\eta$. The size of this effect is
strongly dependent on $L_W$. For thin bubble walls, {\it i.e.}~small
$L_W$, it is negligible, but increases strongly with increasing
$L_W$. As we have fixed the bubble wall velocity and thereby the
diffusion time scale, respectively, the diffusion length scale, the
only length scale in the system that can be different in the parameter
points is the wall thickness. The wall thickness gives the length
along which the bubble profile is changing. The varying Higgs profile
triggers non-zero source terms so that in this region the diffusion
process takes place. The additional massive
particles ($\tau$ and bottom) with their respective source terms can
hence produce more efficiently a left-handed asymmetry for thick bubble
walls resulting in an enhanced BAU compared to the case where only the
top quark contribution is taken into account.
Note finally that the impact of different temperatures
$T_c$ is a slight distortion, respectively, spreading of the
points as can be seen by comparing the left and the right figure.

\section{Conclusions \label{sec:concl}}
In this paper we investigated the question if in principle it is
possible to generate in the C2HDM a baryon asymmetry that is
compatible with the observed value after taking into account all
relevant theoretical and experimental constraints. For this we used
the recent upgrade {\tt BSMPT v2.2} to calculate the BAU in two
different approaches, the {\tt FH} and the {\tt VIA} approach. Our goal
was to investigate differences and similarities of the two methods and
in particular the dependence of the obtained value of $\eta$ on the
various parameters that are relevant for the BAU in order to single
out future directions for upgrades of the implementation and for model
building. \s

We found that both approaches show the same overall behaviour in the
sense that large BAU in the \FH approach also yields large values in
the \VIA approach, with the $\eta$ values computed with {\tt FH} being two to
three orders of magnitude smaller that those obtained from {\tt
  VIA}. The dependence on the wall velocity is mild in the {\tt FH}
approach for small $v_W$ but diverges for velocities near the sound
speed. Recently, however, a re-derivation of the fluid equations
showed that the sound speed barrier can be safely crossed
\cite{Cline:2020jre}. While the results for $\eta$ in the old and new
approach differ by less than 30\% for $v_W=0.1$ the new results of
\cite{Cline:2020jre} will be implemented in future upgrades of our
code. The application of {\tt FH} requires values of $L_W T_c >
1$. In our analysis we found that for parameter points compatible with
the constraints large values of $L_W T_c$ are realized for an overall
light mass spectrum. It turns out, however, that an overall heavier
mass spectrum is advantageous for the BAU. The combination of an
SFOEWPT and the applied constraints on the other hand forbids large mass gaps. These
findings explain why it is difficult to generate a large enough BAU in the
C2HDM compatible with the observed values. Additionally, we need large
CP-violating phases which is in 
contradiction with the strict constraints from the EDM
measurements. As for the impact of $L_W T_c$, $\eta$ shows a similar
behaviour in the {\tt FH} and the {\tt VIA} approach in the region of
large values of $L_W T_c$. Towards smaller values the 
computed $\eta$ with the {\tt FH} method slightly increases. Finally,
we found that the inclusion of additionally the bottom quark besides
the top quark in the transport equations in the {\tt VIA} approach has
a significant increasing effect on $\eta$ 
while the influence of $\tau$ is negligible. Models with additional
fermions might therefore be advantageous for the BAU, a direction that
we investigate in a forthcoming publication. \s

Clearly, the requirement of an SFOEWPT, of a sufficiently large amount of
CP violation and of compatibility with the stringent theoretical and experimental
constraints challenges the generation of a BAU that is
compatible with the observed value. However, the differences in the
results of the calculations from the different methods applied as well
as new insights in the derivation of the bubble wall velocity leave 
room for improvement of the computation of $\eta$. Together with possible
avenues for model building to facilitate an SFOEWPT, to possibly
generate CP violation spontaneously at non-zero temperature thus alleviating the EDM
constraints, or to include new fermions {\it e.g.}~to increase the
obtained value for $\eta$, this gives ample room for further promising
investigations in the context of the dynamical generation of the BAU through
electroweak baryogenesis.

\section*{Acknowledgements}
The research of M.M. was supported by the Deutsche
Forschungsgemeinschaft (DFG, German Research Foundation) under grant
396021762 - TRR 257. P.B. acknowledges financial support  by the Graduiertenkolleg GRK 1694: Elementarteilchenphysik bei h\"ochster Energie und h\"ochster
Pr\"azision. J.M. acknowledges support by the BMBF-Project 05H18VKCC1. 
We are grateful to Stephan Huber for fruitful discussions. We thank
Lisa Biermann for initiating us to reanalyse the renormalization conditions of the
C2HDM resulting in the improved renormalization scheme implemented in
{\tt BSMPT v2.2}. 

\vspace*{1cm}


\end{document}